\documentclass[11pt]{article}
\usepackage[final]{acl}
\usepackage{times}
\usepackage{latexsym}
\newcommand{\pl}{\textsuperscript{+}}
\usepackage{amssymb}
\usepackage{pifont}
\usepackage{booktabs}
\usepackage{makecell}
\usepackage{amsmath}
\usepackage{tabularx}
\usepackage{svg}
\usepackage[most]{tcolorbox}
\tcbuselibrary{skins, breakable}
\usepackage{hyperref}
\usepackage{algorithm}
\usepackage{algorithmic}
\usepackage[T1]{fontenc}
\usepackage[utf8]{inputenc}
\usepackage{listings}
\usepackage[table]{xcolor}
\definecolor{codegreen}{rgb}{0,0.6,0}
\definecolor{codegray}{rgb}{0.5,0.5,0.5}
\definecolor{codepurple}{rgb}{0.58,0,0.82}
\definecolor{backcolour}{rgb}{0.95,0.95,0.92}

\lstdefinestyle{cppstyle}{
    language=C++,
    basicstyle=\ttfamily\footnotesize,
    keywordstyle=\color{blue}\bfseries,
    commentstyle=\color{green!50!black},
    stringstyle=\color{red!60!black},
    breaklines=true,
    frame=single,
    backgroundcolor=\color{white},
    rulecolor=\color{gray!30},
    showstringspaces=false,
    tabsize=2
}

\lstdefinelanguage{json}{
    basicstyle=\ttfamily\footnotesize,
    numbers=left,
    numberstyle=\scriptsize,
    stepnumber=1,
    numbersep=8pt,
    showstringspaces=false,
    breaklines=true,
    frame=single,
    backgroundcolor=\color{gray!5}, 
    literate=
     *{0}{{{\color{blue}0}}}{1}
      {1}{{{\color{blue}1}}}{1}
      {2}{{{\color{blue}2}}}{1}
      {3}{{{\color{blue}3}}}{1}
      {4}{{{\color{blue}4}}}{1}
      {5}{{{\color{blue}5}}}{1}
      {6}{{{\color{blue}6}}}{1}
      {7}{{{\color{blue}7}}}{1}
      {8}{{{\color{blue}8}}}{1}
      {9}{{{\color{blue}9}}}{1}
}

\lstdefinestyle{jsonstyle}{
    language=json,
    stringstyle=\color{red!60!black},
    keywordstyle=\color{blue}\bfseries,
}

\newtcolorbox{promptbox}[1][]{
  colback=gray!5,
  colframe=gray!60,
  title={#1},
  fonttitle=\bfseries,
  sharp corners,
  boxrule=0.5pt,
  left=5pt, right=5pt, top=5pt, bottom=5pt,
  breakable,
  enhanced
}

\usepackage{inconsolata}
\usepackage{graphicx}
\usepackage{pgfplots}
\pgfplotsset{compat=1.17}
\usetikzlibrary{patterns}

\title{CodeHacker: Automated Test Case Generation for Detecting Vulnerabilities in Competitive Programming Solutions}

\author{Jingwei Shi$^{1}$\thanks{Equal contribution.}, Xinxiang Yin$^{2}$\footnotemark[1], Jing Huang$^{3}$\footnotemark[1], Shengyu Tao$^{1}$, Jinman Zhao$^{4}$\thanks{Corresponding author.} \\
  $^1$ Shanghai University of Finance and Economics \quad
  $^2$ Northwestern Polytechnical University \\
  $^3$ Meituan \quad
  $^4$ University of Toronto \\
  \texttt{shijingwei@stu.sufe.edu.cn}
}

\begin{document}
\maketitle
\begin{abstract}
The evaluation of Large Language Models (LLMs) for code generation relies heavily on the quality and robustness of test cases. However, existing benchmarks often lack coverage for subtle corner cases, allowing incorrect solutions to pass. To bridge this gap, we propose \textbf{CodeHacker}, an automated agent framework dedicated to generating targeted adversarial test cases that expose latent vulnerabilities in program submissions. 
Mimicking the \emph{hack} mechanism in competitive programming, CodeHacker employs a multi-strategy approach, including stress testing, anti-hash attacks, and logic-specific targeting to break specific code submissions. 
To ensure the validity and reliability of these attacks, we introduce a Calibration Phase, where the agent iteratively refines its own Validator and Checker via self-generated adversarial probes before evaluating contestant code.
Experiments demonstrate that CodeHacker significantly improves the True Negative Rate (TNR) of existing datasets, effectively filtering out incorrect solutions that were previously accepted. Furthermore, generated adversarial cases prove to be superior training data, boosting the performance of RL-trained models on benchmarks like LiveCodeBench. Our code is available at \url{https://github.com/shi0712/CodeHacker}.
\end{abstract}
\section{Introduction}

Large Language Models~\cite[LLMs,][]{openai2024gpt4technicalreport,deepseekai2025deepseekv3technicalreport} are commonly used for code generations and program reasoning.
Several benchmarks have been developed to evaluate whether LLMs can understand problem constraints correctly, derive correct programs logically, and produce executable implementations~\cite{chen2021evaluatinglargelanguagemodels,Li_2022, wang2025openhandsopenplatformai}.  
One challenge is that the evaluation outcome for each LLM can depend sensitively on the quality and coverage of each test case, raising the questions of how test cases should be constructed and validated properly~\cite{NEURIPS2023_43e9d647}.

A substantial amount of effort has been applied to address this challenge. One line of work focuses on constructing larger and more rigorous benchmarks, thereby reducing bias introduced by sparse test cases.
For example,  LiveCodeBench~\cite{livecodebench} applies stricter evaluation protocols and contamination control, while CodeContests~\cite{codecontest} and CodeContest+~\cite{wang2025codecontests+} target competition-level algorithmic problems with more challenging test settings. Another type of work considers automated test case generation~\cite{taco, hardtest, zhou2025autocode}, attempting to expand test coverage and reach more potential error patterns through input mutation, rule-based construction, or language model generation.  Several studies also rely on human experts to analyze program logic and construct targeted counterexamples or adversarial cases, achieving strong discriminative power at the cost of scalability \citep{tcgbench,wang2025aethercode}. 
However, we argue that these existing automated approaches operate fundamentally as Black-box Fuzzers. 
They rely on probabilistic exploration of the input space, prioritizing statistical coverage to maximize the likelihood of hitting a bug. 
Their evaluation outcomes often still hinge on whether their test cases happen to cover the corresponding failure cases, rather than on whether the model truly understands the problem’s semantics and constraint structure.

Increasing the number of test cases or expanding the input space in a coarse-grained manner is insufficient for boosting the reliability of the evaluation. What ultimately matters is whether test cases can target the specific logical vulnerabilities of a concrete implementation~\citep{li2023taco,hardtest}. In competitive programming practice, constructing a valid counterexample that causes a program to fail (i.e., a successful hack) is often no easier than solving the original problem itself: it requires a deep understanding of algorithmic assumptions, boundary conditions, and complexity constraints, which usually demands fine-grained reasoning about program behavior~\citep{wang2025aethercode}. From this perspective, the ability to generate failure-inducing adversarial inputs reflects not only a model’s understanding of program behavior, but also provides a more discriminative signal for assessing advanced program reasoning capabilities.

While prior work often relies on human experts to generate effective counterexamples (i.e., hacks), and recent advancements have explored multi-agent frameworks for general vulnerability detection~\cite{wang2025vulagent}, we consider the following question: \textit{can such expert-level hacking be automated within a rigorous competitive programming environment?} 
We introduce CodeHacker, a program-centric adversarial framework for competitive programming. 
Unlike methods that rely on heuristic test generation, CodeHacker adopts a Code-Aware Adversarial perspective. CodeHacker treats individual programs as first-class objects and systematically searches for failure-inducing counterexamples, addressing the limitations of existing evaluation approaches in terms of both targeting and scalability. Unlike methods that rely on static test cases or heuristic test generation, our framework adopts a program-centric adversarial evaluation perspective, tightly coupling test construction with the actual execution behavior of the code under evaluation. Our main contributions are summarized as follows:
\begin{enumerate}
    \item We formalize hacking as an adversarial test generation task and design an LLM-driven agent that actively searches for high-value corner cases and logical counterexamples that are difficult to capture with traditional mutation-based or prompt-based generation methods.

    \item By deploying the agent on existing code datasets, we demonstrate substantial improvements in both true negative rate (TNR) and true positive rate (TPR), indicating that the generated adversarial cases meaningfully strengthen evaluation robustness.
    
    \item Building on the generated adversarial cases, we further propose CodeHackerBench, a new evaluation setting designed to better characterize models’ ability to reason about incorrect code and extreme failure scenarios.
\end{enumerate}

\section{Related Work}
\paragraph{Code Benchmark.} The existing code datasets mainly adopt three paradigms.
\emph{Manual} test cases such as MBPP~\cite{mbpp}, HumanEval~\cite{humaneval} and  LiveCodeBench~\cite{livecodebench} are typically handcrafted. Expert-designed cases can better target problem-specific corner cases; manual construction is expensive, hard to automate, and difficult to scale, thus more suitable for small evaluation sets than large training corpora. \emph{Mutation-based} approaches aim to improve coverage by automatically recombining or mutating existing inputs (type-aware input mutation to reduce false positives on MBPP/HumanEval~\cite{NEURIPS2023_43e9d647}, and similar strategies in CodeContests~\cite{codecontest}), but they may violate complex input constraints, potentially introducing invalid tests and raising false negatives. Finally, \emph{LLM-based} generation produces tests conditioned on problem statements~\citep[e.g., Taco][]{taco} and has also been explored in several generators/meta-benchmarks~\citep[e.g., EvalPlus, HardTests, TestCase-Eval, LogiCase, and TCGBench][]{liu2023is,hardtest,testcaseeval,logicase,tcgbench}
; yet LLM outputs are not guaranteed to satisfy intricate constraints and are limited by context/output length, making them less suitable for very large or highly structured instances (e.g., million-node graphs). Furthermore, the evaluation of LLM-generated code has expanded beyond mere functional correctness to encompass execution efficiency~\cite{gong2026traceevaluatingexecutionefficiency}.

\paragraph{Adversarial Data in Competitive Programming.}
Prior work~\cite{hort2025codehacks} introduced \textit{Codehacks}, a dataset constructed by mining historical Codeforces hacks.
Due to the inaccessibility of the original victim submissions via public APIs, this dataset relies on post-hoc matching based on observed execution failures.
Mutation- or randomly generated test approaches expose false negatives but do not model \textit{adversarial intent}~\cite{liu2023is,liu-etal-2025-llm}.

\paragraph{RLHF and RLVR.}
PPO~\cite{ppo} is a standard RLHF optimizer based on on-policy rollouts and value-function estimation. GRPO~\cite{shao2024deepseekmathpushinglimitsmathematical} removes the explicit critic via group-based baselines, and the following works~\cite{dapo, liu2025understanding} further improves training stability. Recent work increasingly adopts RL with verifiable (RLVR) for code generation, often using GRPO-style method~\cite{ekbote2025murphymultiturngrpoself,pennino2025reasoningcodegrpooptimization}. 

\section{Method}
\subsection{Problem Formulation}
\label{sec:problem-formulation}

We consider a Codeforces-style competitive programming environment where an LLM-based agent plays the role of an attacker that attempts to \emph{hack} existing submissions by generating adversarial test cases. {Our use of ``adversarial'' refers to a submission-specific counterexample search under a fixed evaluation protocol. Unlike classical adversarial attacks in machine learning that construct small perturbations, our method searches directly within the valid input space for new test cases that expose latent logical errors.}

Let $\mathcal{P}$ denote the set of problems and, for each problem $p \in \mathcal{P}$, let $\mathcal{X}_p$ be its input space and $\mathcal{Y}_p$ the corresponding output space. 
A contestant submission (program) $s$ is expected to map an input to an output in $\mathcal{Y}_p$, but it may fail during execution. We denote these execution failures as \textbf{Runtime Error (RE)}, \textbf{Time Limit Exceeded (TLE)}, and \textbf{Memory Limit Exceeded (MLE)}. Formally, the submission induces a (partial) mapping:
\[
s : \mathcal{X}_p \to \mathcal{Y}_p \cup \{\text{RE}, \text{TLE}, \text{MLE}\}.
\]

The online judge implements an evaluation function to verify the submission. {Let $J_p(s, T)$ denote the judging verdict of submission $s$ on a test suite $T$, where $J_p(s, T) = \text{AC}$ implies acceptance.}

{We explicitly distinguish the roles of the agent and the environment: the LLM agent operates exclusively on the input space $\mathcal{X}_p$ to synthesize the test input $x$. It does not generate the expected output, preventing potential hallucinations. The corresponding ground truth output $y^*$ is derived externally by the execution environment (running the verified Standard Solution $S_{\text{std}}$). A \textbf{Successful Hack} on a target submission requires three strict conditions:}
\begin{enumerate}
    \item {\textbf{Validity}: The input $x \in \mathcal{X}_{\text{valid}}$ satisfies all problem constraints (verified by our Refined Validator).}
    \item {\textbf{Oracle Confirmation}: The Standard Solution accepts $x$ ($J_p(S_{\text{std}}, \{x\}) = \text{AC}$).}
    \item {\textbf{Target Failure}: The target submission is judged as incorrect ($J_p(s, \{x\}) \neq \text{AC}$).}
\end{enumerate}

{In our framework, we specifically target a set of submissions $S_{\text{target}}$ that exhibit a ground-truth discrepancy between a local benchmark test suite $T_{\text{local}}$ and the official rigorous test suite $T_{\text{official}}$:}
\[
S_{\text{target}} = \left\{ s \in \mathcal{D} \;\middle|\; 
\begin{aligned}
& J_p(s, T_{\text{local}}) = \text{AC} \land {} \\
& J_p(s, T_{\text{official}}) \neq \text{AC}
\end{aligned} 
\right\}
\]

\paragraph{LLM agent as a hacking policy.}
An LLM agent with parameters $\theta$ interacts with a problem--submission pair $(p, s)$ and generates test inputs in order to trigger a non-AC verdict. We model the agent as a (possibly stochastic) policy
\[
\pi_\theta : \mathcal{H}_{p,s} \to \mathcal{X}_p,
\]
where $\mathcal{H}_{p,s}$ denotes the interaction history (including the problem statement, the system feedback, and previously proposed tests). At step $t$, given history $h_t \in \mathcal{H}_{p,s}$, the agent proposes an input
\[
x_t \sim \pi_\theta(\cdot \mid h_t),
\]
which is evaluated by the judge to obtain $v_t = J_p(s, \{x_t\})$. The interaction proceeds for at most $T$ steps or until a non-AC verdict is observed.

\paragraph{Hack success and success rate.}
For a fixed problem--submission pair $(p, s)$ and a given agent $\pi_\theta$, we define the \emph{hack success indicator}
\begin{align}
H(p,s;\pi_\theta)
&= \mathbb{I}\bigl[\exists\, t \le T : J_p(s,\{x_t\}) \neq \text{AC}\bigr].
\end{align}

where $\mathbb{I}[\cdot]$ is the indicator function and the randomness is induced by the policy $\pi_\theta$ (and, if applicable, the environment). Intuitively, the hack succeeds if the agent can find at least one test input within $T$ trials that causes the submission to receive a non-AC verdict.

Given a dataset $\mathcal{D}$ of problem--submission pairs, the overall \emph{hack success rate} (HSR) of the agent is defined as
\[
\mathrm{HSR}(\pi_\theta) 
= \frac{1}{|\mathcal{D}|} 
  \sum_{(p, s) \in \mathcal{D}} 
  \mathbb{E}\bigl[ H(p, s; \pi_\theta) \bigr],
\]
where the expectation is taken over the internal randomness of the LLM agent and, if relevant, the judging system. In our experiments, we empirically approximate $\mathrm{HSR}(\pi_\theta)$ by averaging the observed hack success indicators over all evaluated pairs $(p, s) \in \mathcal{D}$.

\subsection{CodeHacker Agent}

The CodeHacker is an LLM-based agent  responsible for generating adversarial test cases designed to exploit weaknesses in a given program. It interacts with the problem statement and the contestant’s submission to generate inputs that aim to trigger non-AC verdicts such as Wrong Answer (WA), Runtime Error (RE), Time Limit Exceeded (TLE), or Memory Limit Exceeded (MLE).

\begin{figure*}[t]
  \centering
  \includegraphics[width=0.9\textwidth]{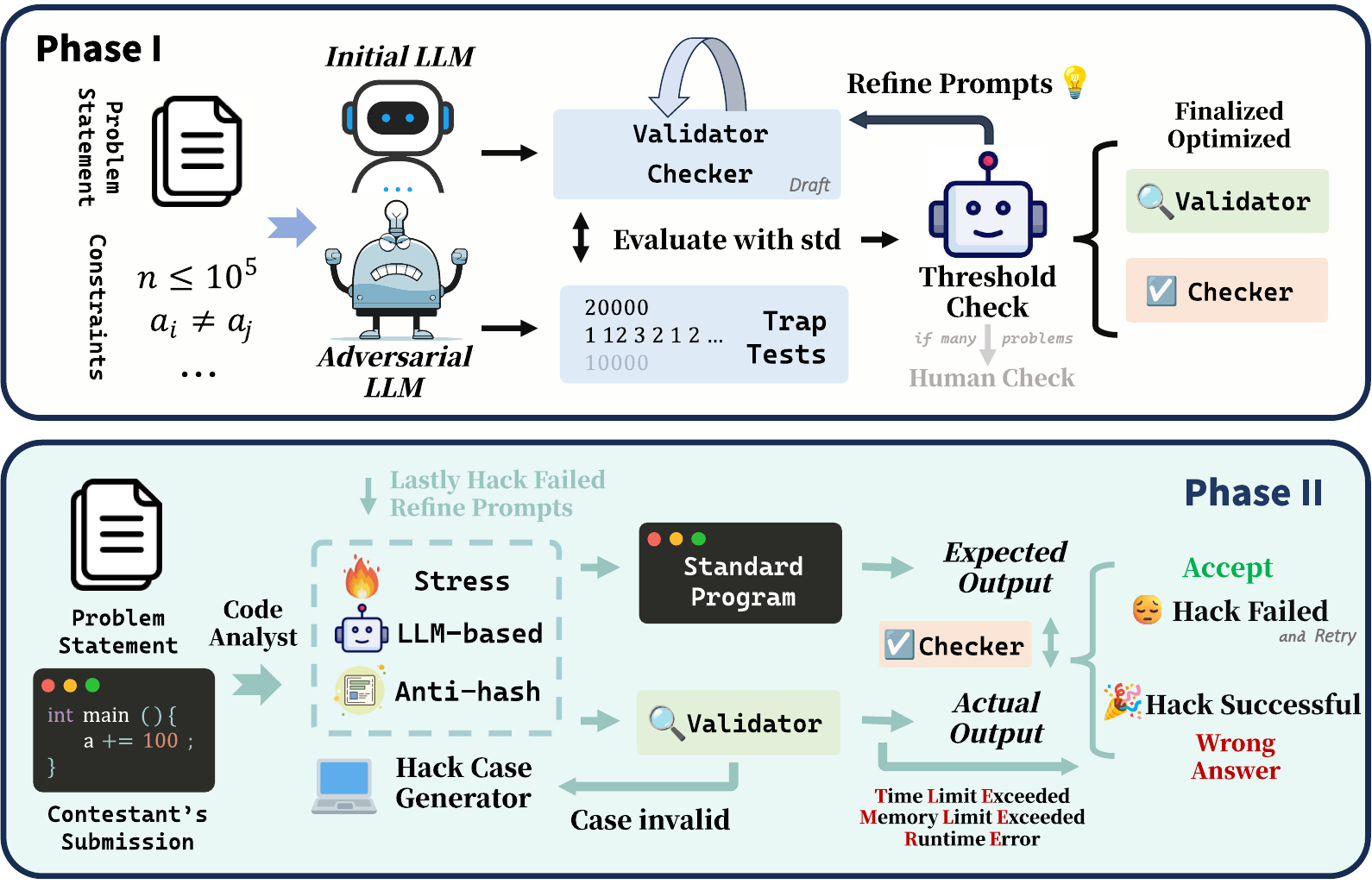}
\caption{The overall architecture of the CodeHacker framework. 
\textbf{Phase I (Evaluation Tool Calibration):} The agent iteratively refines the judging infrastructure to ensure reliability. This process begins by refining the \textbf{Validator} to strictly enforce input constraints, followed by refining the \textbf{Checker} to eliminate false verdicts. 
\textbf{Phase II (Adversarial Case Generation):} Utilizing the calibrated tools, the Code Analyst guides three distinct generation strategies (Stress, LLM-based, and Anti-hash) to synthesize adversarial test cases that expose specific vulnerabilities in the contestant's submission.}
  \label{fig:CodeHacker}
\end{figure*}

\subsection{Phase I: Evaluation Tool Calibration}
\label{sec:calibration}
Before attempting to hack contestant submissions, CodeHacker must ensure its ``ammunition'' (test cases) matches problem constraints and its ``judgment'' (checker) is flawless. We employ an iterative refinement process.

\subsubsection{Validator Refinement}
The Validator is the gatekeeper that ensures test cases lie within the allowed input space $\Phi$. Similar to the Checker, the agent actively attacks the Validator from two directions to identify weaknesses:
\begin{itemize}
    \item \textbf{Bypass Attack ($x_{\text{invalid}}$):} The agent generates an explicitly invalid input $x_{\text{invalid}}$ (e.g., inputs violating constraints like $N > N_{\max}$ or malformed formatting). If the Validator returns \textit{Accepted}, it exposes a False Positive flaw (the validator is too loose and fails to catch illegal inputs).
    \item \textbf{Rejection Attack ($x_{\text{valid}}$):} The agent generates a valid but tricky input $x_{\text{valid}}$ (e.g., legitimate edge cases like $N=1$, values equal to 0, or maximum constraints). If the Validator returns \textit{Rejected}, it exposes a False Negative flaw (the validator is too strict and incorrectly flags valid inputs).
\end{itemize}
If the Validator fails in either case, it is patched to strictly align with the problem definition.
The adversarial refinement process is detailed in \textbf{Algorithm~\ref{algo:validator}} (Appendix~\ref{appx:algos}).

\subsubsection{Checker Refinement}
The Checker verifies the correctness of the program’s output. In the refinement loop, the agent acts as a deceptive adversary, probing for two specific types of vulnerabilities:
\begin{itemize}
    \item \textbf{Deception Attack ($y_{\text{wrong}}$):} The agent constructs an incorrect output $y_{\text{wrong}}$ that mimics the structure of a valid answer (e.g., correct formatting but wrong value, or satisfying only partial constraints). If the Checker returns \textit{Accepted}, it exposes a False Positive flaw (the checker is too loose).
    \item \textbf{Rejection Attack ($y_{\text{true}}$):} The agent identifies a valid output $y_{\text{true}}$ (often an edge case or an alternative valid solution different from the standard reference) that a rigid checker might miss. If the Checker returns \textit{Wrong Answer}, it exposes a False Negative flaw (the checker is too strict).
\end{itemize}
If either attack succeeds, the Checker is flagged as compromised and subsequently updated using the adversarial samples. The complete dual-attack workflow is outlined in \textbf{Algorithm~\ref{algo:checker}} in Appendix~\ref{appx:algos}.

{\paragraph{Anti-Hallucination Pipeline for Checker Update.} To prevent incorrect LLM-generated $y_{\text{true}}$ examples from contaminating the checker update process, we implement a rigorous three-step verification pipeline: (1) \textbf{Small-Scale Inputs}: The adversarial inputs $x$ are restricted to trivial boundary cases where derivation is simple. (2) \textbf{Explicit Reasoning}: The generator LLM must provide a step-by-step Chain-of-Thought to compute $y_{\text{true}}$. (3) \textbf{Cross-Verification}: A separate, independent LLM (LLM-as-a-Judge) audits the generated input, reasoning, and final output against the strict problem constraints. The checker is only updated if it passes this cross-verification.}

\paragraph{Expert Intervention for Complex Verification Logic.}
While our automated refinement pipeline is highly effective, we observed a small fraction of cases where the LLM struggles to generate correct Validators or Checkers.
This typically occurs when the verification logic itself embodies significant algorithmic difficulty. 
For Validators, constraints may require proving the existence of a solution (e.g., ``guarantee the graph has a Hamiltonian path''); for Checkers, verifying a contestant's output can be equivalent to solving a secondary hard problem (e.g., checking graph isomorphism or computing geometric intersections with high precision). 
Since implementing such logic exceeds the typical one-shot capability of LLMs, we employ human expert intervention for these rare instances to manually implement or patch the evaluation tools, ensuring strict adherence to problem semantics. {We clarify that human intervention is required in \textbf{less than 5\%} of cases, primarily concentrated in high-difficulty problems (Codeforces rating $> 2000$). The cost of this intervention is minimal as it is a \textbf{one-time setup cost} per problem. Crucially, human intervention \textbf{does not bias} the reported improvements or LLM evaluation metrics. Experts only intervene in Phase I to ensure the judging system strictly adheres to official semantics. Phase II (Hack Case Generation) remains 100\% autonomous.}

\subsection{Phase II: Adversarial Case Generation}
\label{sec:phase2_execution}

Armed with robust evaluation tools (the calibrated Validator and Checker), CodeHacker proceeds to the execution phase. This phase aims to generate adversarial inputs that specifically trigger non-AC verdicts in the target submission.

\subsubsection{Code Analyst}
\label{sec:code_analyst}

The Code Analyst serves as the strategist. Relying solely on static textual analysis often leads to hallucinations regarding complexity limits or logical edge cases. 
To ensure rigorous vulnerability identification, we equip the Analyst with a \textit{Dual-Execution Interface}, enabling a hybrid analysis workflow:

\begin{itemize}
    \item \textbf{Behavioral Probing (via C++ Sandbox):} 
    The agent treats the target submission as a black box to verify logical hypotheses. 
    For instance, if the agent suspects the code fails on disconnected graphs, it synthesizes a small probe input (e.g., two isolated nodes) and executes the target binary. 
    Observing a crash (RE) or an incorrect output confirms the bug without guessing.
    \vspace{-0.3cm}
    \item \textbf{Precise Calculation (via Python Interpreter):} 
    The agent utilizes Python as a computational engine to verify constraints and mathematical properties. 
    This includes:
    (1) Complexity Verification: Writing scripts to calculate the exact worst-case operation count (e.g., $\sum_{i=1}^{N} \lfloor N/i \rfloor$ for harmonic series) to confirm TLE risks;
    (2) Boundary Checks: Computing whether specific intermediate values (e.g., $C_{2N}^{N}$) exceed the 64-bit integer limit to confirm overflows;
\end{itemize}

By combining behavioral observation with mathematical verification, the Code Analyst formulates a highly reliable hacking plan.

\textbf{Note on Design:} Distinct from the Analyst's minimal logic probes (e.g., $N=5$), the Generator is essential for \textit{weaponizing} these insights into large-scale hacks (e.g., $N=10^5$). By outputting executable C++ code rather than raw data, the Generator bypasses LLM context constraints to produce massive, format-compliant datasets.

\subsubsection{Hack Case Generators}
To maximize the diversity and effectiveness of the attacks, we employ three complementary generation strategies:

\paragraph{Generated by Stress Test:}
This module randomly or systematically explores the boundary of the input space. It focuses on maximizing input size ($N_{max}$) or creating specific structural patterns (e.g., fully connected graphs, skewed trees). While naturally testing asymptotic behavior, this strategy is particularly effective at exposing Runtime Errors (e.g., stack overflows, index out-of-bounds) and Wrong Answers caused by integer overflows or precision issues under extreme constraints.
\vspace{-0.25cm}
\paragraph{Generated by LLM:}
Guided by the Code Analyst's plan, the LLM crafts targeted semantic test cases. This method is versatile and capable of targeting all verdict types. Depending on the vulnerability identified by the Analyst—whether it is an unhandled hack case (WA), a deep recursion risk or segmentation fault (RE), a sub-optimal algorithmic branch (TLE), or inefficient space management (MLE), the LLM synthesizes inputs specifically designed to trigger the corresponding failure mode.
\vspace{-0.25cm}
\paragraph{Generated by Anti-Hash Generator:}
This module targets submissions utilizing hashing algorithms. For fixed-parameter Polynomial Rolling Hashes, we formulate the collision search as a Shortest Vector Problem (SVP) and apply lattice reduction algorithms (e.g., LLL)~\cite{cf_antihash} to deterministically construct collisions. 
Additionally, for scenarios with smaller moduli or non-linear hash functions where lattice reduction is inapplicable, we employ a \textbf{Birthday Attack} strategy. By generating a large pool of random inputs, we exploit the \textit{Birthday Paradox} to identify collisions with high probability in $\mathcal{O}(\sqrt{M})$ complexity, where $M$ is the modulus.Mathematical proofs and implementation details are provided in \textbf{Appendix~\ref{appx:antihash}}.


\subsection{CodeHackerBench}
The \textbf{CodeHackerBench} dataset is constructed by mining judgment discrepancies within the \texttt{CodeContest\pl} dataset. By cross-referencing the ground-truth correctness labels against the verdicts returned by the original (weaker) test cases, we identified approximately 6,000 controversial problem-submission pairs. 

To ensure the absolute reliability of this benchmark, we implemented a Two-Tier Human Verification Protocol covering all critical components (Hack Cases, Validators, and Checkers):
\begin{itemize}
    \item \textbf{Infrastructure Audit (Refined Tools):} We manually reviewed all of the Refined Validators and Checkers modified during the calibration phase. This ensures that our judging logic serves as the gold standard, strictly adhering to the official problem definitions.
    \item \textbf{Hack Case Verification (Data Sampling):} For the Generated Hack Cases, we conducted a random sampling of 600 instances. These cases were inspected by expert competitive programmers, achieving a 100\% validity rate (confirming they adhere to input constraints) and correctly triggering failure modes in the target submissions.
\end{itemize}
\section{Experiments}

\begin{table*}[t]
\centering
\resizebox{0.99\textwidth}{!}{%
\begin{tabular}{lccccc}
\toprule
\multicolumn{1}{c}{\textbf{Dataset}}
& \multicolumn{1}{c}{\textbf{Overall}} 
& \multicolumn{2}{c}{\textbf{Traditional Judge}} 
& \multicolumn{2}{c}{\textbf{Special Judge}} \\
\cmidrule(r){2-2} \cmidrule(r){3-4} \cmidrule(r){5-6}
& \textbf{VPR ($\%\uparrow$)} & \textbf{TPR ($\%\downarrow$)} & \textbf{TNR ($\%\uparrow$)} & \textbf{TPR ($\%\downarrow$)} & \textbf{TNR ($\%\uparrow$)} \\
\midrule
CodeContests \cite{alphacode} & 71.41 & 98.96 & 76.33 & 84.73 & 77.69 \\
HardTests \cite{hardtest}  & 97.32 & 98.33 & 79.25 & 86.56 & 75.37 \\
TACO \cite{li2023taco} & 81.84 & 96.46 & 83.70 & 82.67 & 85.48 \\
CodeContest$^{+}$ \cite{wang2025codecontests+}  & 99.66 & 96.02 & 85.72 & 96.62 & 84.04 \\
\rowcolor{gray!15} 
CodeContest$^{++}$ (Ours) & \textbf{100.00} & \textbf{95.86} & \textbf{96.31} & \textbf{96.38} & \textbf{96.05} \\
\bottomrule
\end{tabular}}

\vspace{1mm}
\begin{minipage}{0.95\linewidth}
    \footnotesize
    \textit{Note:} To ensure evaluation fairness: 
    (1) All test cases were pre-filtered by our \textbf{refined validator} to exclude invalid inputs; 
    (2) Results in the \textbf{Special Judge} columns were evaluated using our \textbf{refined checker} to avoid false verdicts caused by original weak checkers.
\end{minipage}

\caption{Comparison of Overall Validation Pass Rate (VPR) of original datasets, and correctness metrics (TPR/TNR) across 2000 problems with among different datasets.}
\label{tab:dataset-performance}
\end{table*}

\paragraph{Dataset.}
We randomly selected 2000 problems from the \texttt{CodeContest+} dataset, consisting of:
\begin{itemize}
    \item \textbf{1000 traditional judge problems:} These problems rely on the standard online judging systems (Traditional Judge), where the system typically evaluates the solution by comparing the output against the expected results.
    \item \textbf{1000 special judge problems:} These problems use a special judging mechanism (Special Judge), which involves custom evaluation logic. This often includes additional checks such as output formatting, time limits, or other domain-specific rules.
\end{itemize}

In our experiments, we specifically focused on the C++ code of the selected problems as C++ is the most widely used among current competitive programming contests for its high efficiency. Additionally, for RL-based evaluations, we used \textbf{LiveCodeBench}, consisting of 287 problems from AtCoder. The models were evaluated based on the pass@k metrics and \texttt{pass@1}, \texttt{pass@5} indicate the percentage of problems solved by the model within the top-k attempts.

\paragraph{Metrics.} 
We evaluate the quality of our benchmark and agent using four key metrics. Detailed formal definitions and calculation methods are provided in \textbf{Appendix~\ref{appx:metrics}}.
\emph{(1) True Positive Rate (TPR)} and \emph{(2) True Negative Rate (TNR)}: Measure the discriminative power of the test cases against the ground truth labels.
\emph{(3) Validation Pass Rate (VPR)}: Measures the proportion of test cases in a dataset that satisfy the strict problem constraints (validity).
Additionally, we report the \emph{Hack Success Rate (HSR)} to quantify agent performance.

\paragraph{Implementation Details.} 
We randomly selected 2,000 problems from \texttt{CodeContest\pl}. Crucially, to ensure verdict fidelity, we strictly replicated the official Codeforces Windows-based toolchain (MinGW)~\cite{codeforces_cmd}. For the generation modules, we set the sampling temperature to $0.7$, preventing the agent from collapsing into repetitive patterns and encouraging a broader exploration of the input space.
For Reinforcement Learning, we train Qwen3-4B using the DAPO algorithm~\cite{yu2025dapo}. 
Complete implementation details, including exact compiler flags, infrastructure setups, and RL hyperparameters, are detailed in \textbf{Appendix~\ref{appx:compilation}} and \textbf{Appendix~\ref{appx:rl_details}}.

\subsection{Main Results}
\paragraph{Adversarial hacking acts as a discriminator for advanced reasoning.}
Table~\ref{tab:hack-success-rate} shows that DeepSeek V3.2 achieves the highest HSR of 64.83\%, outperforming GPT-5-Mini (51.40\%) and Gemini-3.0 (35.65\%). Crucially, we observe a dramatic performance gap when explicit reasoning is disabled: DeepSeek V3.2's success rate plummets to 23.76\% (a 2.7$\times$ decline) in non-thinking mode. This substantial performance gap between reasoning and non-reasoning models underscores a critical finding: \textbf{Hacking is a Reasoning Task}. 
Complex adversarial inputs such as specific tree topologies to maximize recursion depth or mathematically precise sequences cannot be retrieved from memory or guessed. 
They require the model to perform \textit{constructive reasoning}: deriving a generative algorithm that satisfies multiple entangled constraints. 
This explains why standard LLMs fail to generate valid complex hacks, whereas reasoning-enhanced models excel.

\begin{table}[]
\centering
\resizebox{0.97\linewidth}{!}{%
\begin{tabular}{l c c}
\toprule
\textbf{Backbone Model} & \textbf{HSR (\%)} & \textbf{Avg. \#T} \\
\midrule

\multicolumn{3}{l}{\textbf{Reasoning Models}} \\
DeepSeek V3.2 & \textbf{64.83} & 1.24 \\
GPT-5-Mini & 51.40 & 1.35 \\
Qwen3-Next-80B-A3B & 38.37 & 1.58 \\
Gemini-3.0-Flash-Preview & 35.65 & 1.62 \\
Qwen3-32B & 23.64 & 1.89 \\

\midrule

\multicolumn{3}{l}{\textbf{Non-Reasoning Models}} \\
DeepSeek V3.2$^*$ & \textbf{23.76} & 2.45 \\
Qwen3-Next-80B-A3B$^*$ & 20.06 & 2.68 \\
Qwen3-32B$^*$ & 10.00 & 2.91 \\
\bottomrule
\end{tabular}}
\caption{Hack performance across different models. \textbf{HSR}: Hack Success Rate. \textbf{Avg. \#T}: The average number of refinement turns required to successfully hack a submission by directly LLM generated (lower is better, max 5). The symbol * indicates non-thinking mode for hybrid-reasoning models.}
\label{tab:hack-success-rate}
\vspace{-0.5cm}
\end{table}

{\paragraph{Performance Degradation as Metric Correction.} To quantify the "inflation" in current evaluations, we evaluated SOTA models on CodeHackerBench. As shown in Table~\ref{tab:metric-correction}, Pass@1 scores drop across the board. We argue this is a \textbf{correction of metric inflation} rather than a capability loss. Standard benchmarks allow flawed solutions ("False Positives") to pass due to weak test coverage. CodeHackerBench filters these out. Notably, we observe a rank reversal: while Gemini-3.0-Flash initially outperformed GPT-5-Mini on original tests, our rigorous adversarial evaluation reveals that GPT-5-Mini is actually more robust.}

\begin{table}[h]
\centering
\resizebox{0.99\linewidth}{!}{%
\begin{tabular}{l c c c}
\toprule
\textbf{Model} & \textbf{Pass@1} & \textbf{Adjusted Pass@1} & $\Delta$ \\
\midrule
DeepSeek V3.2 & 83.3\% & 81.6\% & -1.7\% \\
GPT-5-Mini & 78.4\% & 76.7\% & -1.7\% \\
Gemini-3.0-Flash & 78.6\% & 76.5\% & -2.1\% \\
DeepSeek V3.2* & 61.9\% & 59.7\% & -2.2\% \\
Qwen3-32B* & 51.0\% & 47.5\% & -3.5\% \\
\bottomrule
\end{tabular}}
\caption{{Pass@1 Performance Correction on CodeHackerBench. The drop quantifies the proportion of "lucky" submissions that contained latent bugs but passed the original weak tests.The symbol * indicates non-thinking mode for hybrid-reasoning models.}}
\label{tab:metric-correction}
\vspace{-0.3cm}
\end{table}

\paragraph{Adversarial hacking corrects metric inflation and enhances evaluation rigor.}
To accurately assess the impact of adversarial test cases, we constructed a new benchmark variant, \textbf{CodeContests\pl\pl}.
We posit that under the strict premises of input validity and verdict correctness (guaranteed by our refined validator and checker), the hallmark of a rigorous benchmark is a significantly higher TNR coupled with a moderately lower TPR.
As shown in Table~\ref{tab:dataset-performance}, unlike baselines with inflated TPRs ($\approx$99\%) that often mask subtle flaws, CodeHacker achieves a superior TNR (e.g., \textbf{96.05\%} on Special Judge problems).
This performance confirms that our framework successfully corrects historical misclassifications: the observed drop in TPR is not a degradation but a correction of \emph{inflated} metrics, exposing latent bugs in previously ``Accepted'' solutions rather than incorrectly rejecting valid code.

\paragraph{Adversarial data improves RL efficiency and generalization.}
Table~\ref{fig:rl_pass5_performance} shows that the model RL-trained on the augmented \texttt{CodeContests\pl\pl} consistently outperforms the baseline trained on standard data  (see Appendix~\ref{appx:detailed_rl_results} for detailed numerical breakdowns). This indicates that high-quality adversarial inputs serve as dense reward signals during reinforcement learning, forcing the policy to optimize for strict boundary condition handling and deep logical robustness rather than simple pattern matching. Crucially, the performance gains on the out-of-distribution LiveCodeBench confirm that learning from hack cases fosters genuine algorithmic reasoning capabilities rather than overfitting to the training set. 

\begin{figure}[t!]
\centering
\begin{tikzpicture}
    \begin{axis}[
        ybar,
        bar width=14pt,
        width=0.98\linewidth,
        height=6cm,
        enlarge x limits=0.25,
        legend style={at={(0.5,1.1)}, anchor=south, legend columns=-1, draw=none, fill=none, font=\small},
        ylabel={Pass@5 Acc (\%) on LiveCodeBench},
        symbolic x coords={Easy, Medium, Hard},
        xtick=data,
        nodes near coords,
        nodes near coords style={font=\tiny, /pgf/number format/.cd, fixed, precision=1},
        ymin=0, ymax=115,
        ymajorgrids=true,
        grid style=dashed,
        cycle list={
            {fill=gray!40, draw=none},        
            {fill=blue!50, draw=none},        
            {fill=red!60!black, draw=none}    
        }
    ]
    \addplot coordinates {(Easy, 97.01) (Medium, 73.53) (Hard, 25.00)};
    \addplot coordinates {(Easy, 97.01) (Medium, 75.00) (Hard, 26.97)};
    \addplot coordinates {(Easy, 98.51) (Medium, 80.88) (Hard, 27.63)};
    \legend{Base Model, RL w/ CC+, \textbf{RL w/ Ours}}
    \end{axis}
\end{tikzpicture}
\caption{Pass@5 performance comparison on LiveCodeBench. Models trained with adversarial data show consistent improvements. The model trained on our augmented subset achieves the highest accuracy across all difficulty levels.}
\label{fig:rl_pass5_performance}
\vspace{-0.5cm}
\end{figure}

\subsection{Ablation Study}
To rigorously evaluate the contributions of our framework, we conduct a two-fold ablation study. First, we analyze the impact of individual agent modules on the Hack Success Rate (HSR), determining which components are essential for generating valid attacks. Second, we examine the Sanitation Pipeline step-by-step to understand how each refinement stage improves the reliability (TNR and TPR) of the final benchmark.

\paragraph{Impact of Agent Components}
We investigate the necessity of the Code Analyst, Refinement Loop, and Stress Test modules by removing them one by one from the CodeHacker agent. As shown in Table~\ref{tab:ablation-components}, the full CodeHacker agent achieves the highest success rate (51.40\% using GPT-5-mini).

\begin{table}[h]
\centering
\resizebox{0.99\linewidth}{!}{%
\begin{tabular}{lc}
\toprule
\textbf{Variant} & \textbf{Hack Success Rate (\%$\uparrow$)} \\
\midrule
\textbf{CodeHacker (Full)} & \textbf{51.40} \\
\quad w/o Code Analyst & 49.86 \\
\quad w/o Stress Test & 50.12 \\
\quad w/o Anti-Hash Generator & 51.00 \\
\quad w/o Refinement Loop & 46.60 \\
\bottomrule
\end{tabular}}
\caption{Ablation study of CodeHacker agent components.}
\label{tab:ablation-components}
\vspace{-0.3cm}
\end{table}

The Refinement Loop proves to be the most critical component; its removal causes a significant performance drop ($51.40\% \to 46.60\%$). This highlights the difficulty of generating strictly valid adversarial cases in a single pass—iterative feedback is essential for correcting invalid inputs. The Code Analyst also plays a vital role ($49.86\% \to 51.40\%$) by guiding the generator toward logical vulnerabilities rather than blind guessing. The Stress Test module provides a minor but necessary boost ($50.12\% \to 51.40\%$) by covering asymptotic complexity failures that semantic analysis might miss.

It is worth noting that while the Anti-Hash Generator was triggered in only a small fraction of cases (approximately 0.4\% of submissions using rolling hashes), it achieved a 100\% success rate in breaking these solutions. This confirms that while niche, domain-specific adversarial modules are essential for achieving perfect coverage against algorithmic shortcuts.

\paragraph{Impact of Dataset Enhancement Pipeline}
We further analyze how our infrastructure improvements—specifically the tool calibration (Validator/Checker) and the adversarial data augmentation (Hack Cases) affect evaluation reliability. To isolate the impact of judging logic, we conduct this ablation study specifically on the 1000 problems requiring Special Judges. Table~\ref{tab:ablation-stepwise} details the cumulative effect of the refined validator, refined checker, and the addition of hack cases.

\begin{table}[h]
\centering
\resizebox{0.99\linewidth}{!}{%
\begin{tabular}{l c c}
\toprule
\textbf{Configuration} & \textbf{TNR (\%$\uparrow$)} & \textbf{TPR (\%)} \\
\midrule
Baseline (CodeContest$^{+}$) & 82.18 & 95.34 \\
\quad + Refined Validator & 82.08 & 95.50 \\
\quad + Refined Checker & 84.04 & 96.62 \\
\rowcolor{gray!15}
\quad + Hack Cases (Ours) & \textbf{96.05} & 96.38 \\
\bottomrule
\end{tabular}}
\caption{Stepwise analysis of the dataset enhancement pipeline on Special Judge problems. Adding adversarial Hack Cases provides the decisive boost to TNR.}
\label{tab:ablation-stepwise}
\vspace{-0.3cm}
\end{table}

An interesting phenomenon occurs when introducing the refined validator: the TNR slightly drops from 82.18\% to 82.08\%. This reveals a \textit{masking effect} in the baseline, where invalid inputs previously caused incorrect solutions to fail, inadvertently counting them as correctly rejected. By filtering these invalid inputs, the refined validator exposes the baseline's true, weaker discriminatory power.

While the refined checker improves verdict precision (raising TPR to 96.62\%), the most substantial improvement in robustness comes from the augmentation with hack cases. This step leaps the TNR to 96.05\%, confirming that standard test cases lack the coverage to catch subtle logical errors, and that adversarial generation is non-negotiable for rigorous evaluation.
\section{Conclusion}
In this work, we introduced \textbf{CodeHacker}, an autonomous framework that iteratively refines competitive programming evaluations via a novel Self-Hacking mechanism. By correcting legacy judging flaws and augmenting the test cases with targeted adversarial inputs, our augmented subset achieves highest True Negative Rate (TNR), offering a significantly more trustworthy standard than existing benchmarks. Furthermore, we presented \textbf{CodeHackerBench}, a benchmark designed to evaluate the \textit{adversarial reasoning} capabilities of LLMs. Our experiments demonstrate that the ability to generate valid hacks serves as a strong differentiator for advanced reasoning models, mirroring the cognitive demands of rigorous algorithm design. We hope this work establishes adversarial generation as a critical aspect for evaluating general algorithmic intelligence. Beyond academic evaluation, we envision CodeHacker as a vital component of the next-generation competitive programming infrastructure, assisting both problem setters in quality assurance and contestants in personalized training. {Furthermore, our framework is conceptually transferable to real-world issue-solving benchmarks such as SWE-bench and broader repository-level evaluations~\cite{li2026laboratory,lian2026sweagilesoftwareagentframework}. The core principle-systematic regression-test augmentation that probes untested boundary conditions to challenge overestimated correctness—can be applied to broader software engineering tasks to improve evaluation rigor in real-world applications.}

\clearpage
\newpage
\section*{Ethic Statement}
\label{sec:ethics}

We emphasize two critical ethical principles regarding the research and application of automated hacking agents, strictly adhering to the community standards of the competitive programming ecosystem:

\paragraph{Prohibition of Unauthorized Data Scraping.}
While platforms like Codeforces and QOJ publicly display hack attempts, they explicitly prohibit the use of automated web crawlers or scrapers to harvest such data in bulk. Unauthorized scraping violates the Terms of Service of these platforms and places an unsustainable load on community-maintained servers. Our research strictly respects these prohibitions; we urge future researchers to avoid unauthorized data harvesting and to rely solely on officially released datasets or compliant data access methods to preserve platform integrity.

\paragraph{Mandatory Local Evaluation.}
It is fundamentally impermissible to evaluate AI agents by submitting generated test cases to public Online Judges. Using public judging queues for automated testing constitutes a Denial of Service (DoS) risk and degrades the service quality for human contestants. Therefore, it is mandatory that all adversarial evaluations be conducted in a local, sandboxed environment. CodeHacker is specifically designed with a Self-Hacking loop to operate entirely offline, ensuring zero interference with live judging infrastructure.

\section*{Limitations}
While our approach has shown promising results, there are several limitations that need to be addressed in future work. While the CodeHacker agent is capable of generating valuable test cases, it is not infallible. There may still be certain edge cases or subtle algorithmic flaws that the agent fails to identify. Expanding the agent’s capability to detect more complex failures, especially those related to performance (e.g., time or memory limitations), remains an open challenge.
Besides, our current evaluation focuses on a subset of programming languages and problem domains. Expanding the CodeHackerBench benchmark to include additional languages and a broader set of problem types will provide a more comprehensive assessment of the approach’s generalizability. As competitive programming and algorithmic challenges evolve, we plan to adapt our framework to ensure it stays up-to-date with the latest trends and challenges in the field.

\bibliography{custom}
\appendix

\section{Terms and Definitions}
To ensure clarity and consistency throughout this paper, we provide formal definitions for standard terms used in the competitive programming ecosystem. The evaluation of a solution typically involves a rigorous interaction between the contestant's code and the Online Judge (OJ) system's components. Understanding the precise roles of the \textit{Validator}, \textit{Checker}, and the mechanics of a \textit{Hack} is essential for grasping the workflow of our CodeHacker framework. Table~\ref{tab:terms} summarizes these key definitions and their specific contexts within this work.
\newcommand{\cterm}[2]{%
  \shortstack[l]{\textbf{#1}\\\textit{#2}}%
}

\begin{table*}[htbp]
\normalsize
\centering
\caption{Common Terms in Competitive Programming}
\renewcommand{\arraystretch}{1.2}
\begin{tabularx}{\textwidth}{@{}>{\raggedright\arraybackslash}p{4.5cm} >{\raggedright\arraybackslash}X@{}}
\toprule
\textbf{Term} & \textbf{Description} \\ \midrule

\textbf{Submission} &
In programming competitions, a submission is a program submitted by a contestant.
This submission is evaluated by a judging system, resulting in a verdict such as
\textit{Accepted (AC)}, \textit{Wrong Answer (WA)}, \textit{Time Limit Exceeded (TLE)},
\textit{Runtime Error (RE)}, or \textit{Compile Error (CE)}, among others. \\[0.4em]

\textbf{Test case} &
A test case is used to check whether the participant’s submission is correct.
It usually consists of input data and the corresponding correct output (reference answer). \\[0.4em]

\textbf{Test input} &
The input data of a test case will be fed into the contestant’s program.
The program’s output will then be compared with the reference output to determine correctness. \\[0.4em]

\textbf{Test output}&
The output data of a test case is the correct answer corresponding to the input data.
For problems with multiple correct solutions, the reference output typically represents
one of the possible correct answers. \\[0.4em]

\textbf{Validator} &
A validator is a program used to check whether the input data satisfies the problem
constraints and format. \\[0.4em]

\textbf{Generator}&
A generator is a program used to produce test input automatically, especially when
the input data is large, random, or complex and cannot be easily handcrafted. \\[0.4em]

\textbf{Checker}&
A checker is a program used to determine if a contestant’s output is correct.
Usually, it simply compares the contestant’s output with the expected output, but for
problems with multiple solutions (e.g., floating-point answers or different valid orderings),
it may include more complex judging logic. \\[0.4em]

\textbf{Hack} &
In \textit{Codeforces}, a hack is a special mechanism available during the hacking phase
of some rounds. Participants can test other contestants’ solutions by providing a test case
that causes them to fail. If a hack succeeds, the hacker gains additional points, and the
hacked participant may lose points or their solution becomes marked as \textit{hacked}.
Hacks help ensure solution robustness and encourage writing well-tested code. \\

\bottomrule
\label{tab:terms}
\end{tabularx}
\end{table*}

\section{Refinement Algorithms}
\label{appx:algos}

We provide the detailed pseudocode for the iterative refinement of the Validator and Checker below.

\begin{algorithm}[h]
\small
\caption{Iterative LLM-based Refinement of Validator}
\label{algo:validator}
\begin{algorithmic}[1]
\REQUIRE Problem constraints $\Phi$, validation prompt $P_{\text{val}}$, hacking prompt $P_{\text{val-hack}}$, LLM $\mathcal{M}$, max failures $K$
\ENSURE Validator $V$
\STATE $V \leftarrow \mathcal{M}(P_{\text{val}}, \Phi)$
\STATE $k_{\text{fail}} \leftarrow 0$
\WHILE{$k_{\text{fail}} < K$}
    \STATE $(x_{\text{valid}}, x_{\text{invalid}}) \leftarrow \mathcal{M}(P_{\text{val-hack}}, \Phi, V)$
    \STATE $\text{is\_FP} \leftarrow (V(x_{\text{invalid}}) == \text{Accepted})$
    \STATE $\text{is\_FN} \leftarrow (V(x_{\text{valid}}) \neq \text{Accepted})$
    \IF{$\text{is\_FP} \lor \text{is\_FN}$}
        \STATE $k_{\text{fail}} \leftarrow 0$
        \STATE Update $P_{\text{val}}$ with failure cases $x_{\text{invalid}}$ and $x_{\text{valid}}$
        \STATE $V \leftarrow \mathcal{M}(P_{\text{val}}, \Phi)$
    \ELSE
        \STATE $k_{\text{fail}} \leftarrow k_{\text{fail}} + 1$
    \ENDIF
\ENDWHILE
\STATE \textbf{return} $V$
\end{algorithmic}
\end{algorithm}

\begin{algorithm}[h]
\small
\caption{Iterative LLM-based Refinement of Checker}
\label{algo:checker}
\begin{algorithmic}[1]
\REQUIRE Problem spec $\Phi$, checking prompt $P_{\text{c}}$, hacking prompt $P_{\text{c-hack}}$, oracle $\mathcal{O}$, LLM $\mathcal{M}$, max failures $K$
\ENSURE Checker $C$
\STATE $C \leftarrow \mathcal{M}(P_{\text{c}}, \Phi)$
\STATE $k_{\text{fail}} \leftarrow 0$
\WHILE{$k_{\text{fail}} < K$}
    \STATE $(x_{\text{cand}}, y_{\text{wrong}}, y_{\text{true}}) \leftarrow \mathcal{M}(P_{\text{c-hack}}, \Phi, C)$
    \STATE $is\_FP \leftarrow (C(x_{\text{cand}}, y_{\text{wrong}}) == \text{Accepted})$
    \STATE $is\_FN \leftarrow (C(x_{\text{cand}}, y_{\text{true}}) \neq \text{Accepted})$
    \IF{$is\_FP$ \OR $is\_FN$}
        \STATE $k_{\text{fail}} \leftarrow 0$
        \STATE $y_{\text{gt}} \leftarrow \mathcal{O}(\Phi, x_{\text{cand}})$ 
        \STATE Update $P_{\text{c}}$ with failure case $(x_{\text{cand}}, y_{\text{wrong}}, y_{\text{true}},y_{\text{gt}})$
\STATE $C \leftarrow \mathcal{M}(P_{\text{c}}, \Phi)$
    \ELSE
        \STATE $k_{\text{fail}} \leftarrow k_{\text{fail}} + 1$
    \ENDIF
\ENDWHILE
\STATE \textbf{return} $C$
\end{algorithmic}
\end{algorithm}

\section{Codeforces C++ Compilation Environments}
\label{appx:compilation}

To ensure full reproducibility and consistency with competitive-programming settings, we faithfully replicated all official Codeforces C++ compiler configurations. 
Contrary to common Linux-based setups, Codeforces officially employs a Windows-based judging infrastructure. Consequently, we strictly utilized the official flags (e.g., \texttt{-Wl,--stack=268435456}) to prevent discrepancies in TLE or RE verdicts.

\begin{table*}[h!]
\centering
\small
\begin{tabular}{lll p{0.55\linewidth}}
\toprule
\textbf{Compiler} & \textbf{Arch} & \textbf{Version} & \textbf{Compilation Command} \\
\midrule
GNU G++14 & 32-bit & 6.4.0 &
\texttt{-static -DONLINE\_JUDGE -Wl,--stack=268435456 -O2 -std=c++14} \\
GNU G++17 & 32-bit & 7.3.0 &
\texttt{-static -DONLINE\_JUDGE -Wl,--stack=268435456 -O2 -std=c++17} \\
GNU G++20 & 64-bit & 11.2.0 (WinLibs) &
\texttt{-Wall -Wextra -Wconversion -static -DONLINE\_JUDGE -Wl,--stack=268435456 -O2 -std=c++20} \\
GNU G++23 & 64-bit & 14.2 (MSYS2) &
\texttt{-Wall -Wextra -Wconversion -static -DONLINE\_JUDGE -Wl,--stack=268435456 -O2 -std=c++23 -lstdc++exp} \\
\bottomrule
\end{tabular}
\caption{C++ compilation environments reproduced from the official Codeforces judge configuration.}
\label{tab:cf-env}
\end{table*}

\section{RL Training Hyperparameters}
\label{appx:rl_details}

We employed the \textbf{DAPO} algorithm~\cite{yu2025dapo} to train the Qwen3-4B base model. 
To accommodate the lengthy context often required in competitive programming (e.g., long problem statements and complex logic), we set the maximum generation length to 24,000 tokens.
The training was conducted with a global batch size of 32 and an actor learning rate of $5 \times 10^{-7}$. During the exploration phase, we sampled $N=8$ solutions per prompt. We adopted a dual clipping strategy with $\epsilon_{\text{low}}=0.2$ and $\epsilon_{\text{high}}=0.3$. 
The reward is rule-based: the model receives a reward of $+1$ if the generated code successfully passes all test cases, and $0$ otherwise.

We provide the detailed hyperparameters used for the DAPO training in Table~\ref{tab:rl_params}. All experiments were conducted on a cluster of NVIDIA H100 GPUs.
\begin{table}[h]
\centering
\small
\renewcommand{\arraystretch}{1.2}
\begin{tabular}{l c}
\toprule
\textbf{Hyperparameter} & \textbf{Value} \\
\midrule
Algorithm & DAPO \\
Base Model & Qwen3-4B \\
Train Batch Size & 32 \\
Samples per Prompt ($N$) & 8 \\
Actor Learning Rate & $5 \times 10^{-7}$ \\
Max Generation Length & 24,000 \\
Clip Range ($\epsilon_{\text{low}}, \epsilon_{\text{high}}$) & 0.2, 0.3 \\
\bottomrule
\end{tabular}
\caption{Hyperparameters for Reinforcement Learning (DAPO).}
\label{tab:rl_params}
\end{table}

\section{Metric Definitions and Calculations}
\label{appx:metrics}

To ensure the reproducibility and clarity of our evaluation, we provide the formal definitions for all metrics used in the experimental section.

Let $\mathcal{D}$ be the dataset of problem-submission pairs. For a specific submission $s \in \mathcal{D}$, let $J_{\text{orig}}(s)$ denote the verdict provided by the original dataset (ground truth), and $J_{\text{new}}(s)$ denote the verdict obtained using our augmented test cases (CodeHacker generated cases).

We define the binary correctness function $V(s)$ as:
\[
V(\text{verdict}) = \begin{cases} 
1 (\text{Correct}) & \text{if verdict is \textbf{Accepted}} \\
0 (\text{Incorrect}) & \text{otherwise.}
\end{cases}
\]

\subsection{True Positive Rate (TPR) and True Negative Rate (TNR)}
These metrics evaluate the discriminative power of the test case with respect to the original ground truth labels.

\paragraph{True Positive Rate (TPR):} The proportion of solutions originally marked as correct that are also accepted by the new test case.
\[
\text{TPR} = \frac{\sum_{s \in \mathcal{D}_{\text{pos}}} \mathbb{I}[V(J_{\text{new}}(s)) = 1]}{|\mathcal{D}_{\text{pos}}|}
\]
where $\mathcal{D}_{\text{pos}} = \{s \in \mathcal{D} \mid V(J_{\text{orig}}(s)) = 1\}$. A TPR significantly lower than 100\% implies that the new test case has successfully hacked solutions that were previously thought to be correct (exposing False Positives in the original data).

\paragraph{True Negative Rate (TNR):} The proportion of solutions originally marked as incorrect that are correctly rejected by the new test case.
\[
\text{TNR} = \frac{\sum_{s \in \mathcal{D}_{\text{neg}}} \mathbb{I}[V(J_{\text{new}}(s)) = 0]}{|\mathcal{D}_{\text{neg}}|}
\]
where $\mathcal{D}_{\text{neg}} = \{s \in \mathcal{D} \mid V(J_{\text{orig}}(s)) = 0\}$. A low TNR indicates that a test case is too weak, allowing incorrect solutions to pass (False Positives).

\subsection{Validation Pass Rate (VPR)}
VPR measures the quality of the test cases themselves, specifically ensuring they adhere to problem constraints.
Let $T_{\text{orig}}$ be the set of all test cases in the original dataset. Let $\mathcal{V}(t)$ be our \textbf{Refined Validator}, which returns 1 if an input $t$ satisfies all problem constraints and 0 otherwise.

\[
\text{VPR} = \frac{\sum_{t \in T_{\text{orig}}} \mathcal{V}(t)}{|T_{\text{orig}}|} \times 100\%
\]
A VPR less than 100\% indicates that the original dataset contained invalid test cases that were subsequently filtered out in our benchmark.

\subsection{Hack Success Rate (HSR)}
HSR measures the agent's ability to trigger a failure in a submission. For a specific set of target submissions $S_{\text{target}}$:

\[
\text{HSR} = \frac{\sum_{s \in S_{\text{target}}} \mathbb{I}[\exists x \in X_{\text{gen}} : J(s, x) \neq \text{AC}]}{|S_{\text{target}}|}
\]
where $X_{\text{gen}}$ is the set of adversarial inputs generated by the agent for submission $s$.

\section{Data Decontamination and Benchmark Independence}
\label{appx:decontamination}

To ensure the validity of our reinforcement learning results and strictly prevent data leakage, we implemented a \textbf{Source-Based Decontamination Protocol}.

\paragraph{Source Separation.}
Our training and evaluation datasets are derived from disjoint algorithmic contest platforms, ensuring structural independence:
\begin{itemize}
    \item \textbf{Training Set (\texttt{CodeContests\pl\pl}):} The augmented training data is exclusively sourced from \textbf{Codeforces} problems.
    \item \textbf{Evaluation Set (LiveCodeBench):} For the evaluation of RL-trained models, we specifically utilized the subset of LiveCodeBench consisting entirely of \textbf{AtCoder} problems.
\end{itemize}
Since Codeforces and AtCoder are distinct platforms with unique problem sets, checking systems, and editorial styles, there is effectively \textbf{zero overlap} between our training and evaluation data. This physical separation guarantees that the performance improvements observed in Table~\ref{tab:passk_difficulty_lcb_appendix} are due to the generalization of adversarial reasoning capabilities rather than memorization of specific problem patterns.

\begin{table*}[h]
\centering
\resizebox{0.99\textwidth}{!}{%
\begin{tabular}{lcc|cc|cc}
\toprule
\textbf{Training Stage} & \textbf{Easy pass@1} & \textbf{Easy pass@5} & \textbf{Medium pass@1} & \textbf{Medium pass@5} & \textbf{Hard pass@1} & \textbf{Hard pass@5} \\
\midrule
Base Model (Qwen3-4B) & 91.04 & 97.01 & 63.24 & 73.53 & 11.84 & 25.00 \\
RL w/ CodeContests\pl & 94.03 & 97.01 & 58.82 & 75.00 & 15.13 & 26.97 \\
\rowcolor{gray!15}
RL w/ CodeContests\pl\pl & \textbf{94.03} & \textbf{98.51} & \textbf{66.18} & \textbf{80.88} & \textbf{17.11} & \textbf{27.63} \\
\bottomrule
\end{tabular}}
\caption{Detailed Pass@$k$ performance on LiveCodeBench across different difficulty levels. Comparing the Base Model against versions trained via RL using standard vs. adversarial datasets.}
\label{tab:passk_difficulty_lcb_appendix}
\end{table*}

\section{Detailed Reinforcement Learning Results}
\label{appx:detailed_rl_results}

In this section, we provide the comprehensive numerical results for the Reinforcement Learning evaluation on LiveCodeBench. Table~\ref{tab:passk_difficulty_lcb_appendix} details the pass@1 and pass@5 accuracy across three difficulty levels (Easy, Medium, Hard).

\section{Mathematical Derivation of Anti-Hash Generator}
\label{appx:antihash}

In this section, we detail the construction used to generate hash collisions for polynomial rolling hashes with fixed moduli $\{p_i\}_{i=0}^{n-1}$ and bases $\{q_i\}_{i=0}^{n-1}$.

\subsection{Problem Definition}
The hash function for a string $a$ of length $L$ is defined as:
\[
h_i(a) = \left( \sum_{j=0}^{L-1} a_j q_i^j \right) \pmod{p_i}
\]
Our goal is to find two distinct strings $a$ and $b$ such that $h_i(a) = h_i(b)$ for all $i$. This is equivalent to finding a difference array $d = a - b$ satisfying:
\begin{enumerate}
    \item $\sum_{j=0}^{L-1} d_j q_i^j \equiv 0 \pmod{p_i}$ for all $i$.
    \item $|d_j| < 26$ (assuming lowercase English letters) for all $j$, and $d \neq 0$.
\end{enumerate}

\subsection{Lattice Construction}
We formulate this as a Shortest Vector Problem (SVP). We construct a lattice basis matrix $M$ of size $(n + L) \times (n + L)$. To prioritize satisfying the modular equations (i.e., forcing the remainder to be 0), we introduce a large weight factor $\lambda$.

The matrix $M$ is defined as a block matrix:
\[
M = 
\begin{bmatrix}
\lambda Q & I \\
\lambda P & 0 
\end{bmatrix}
\]
where:
\begin{itemize}
    \item $Q$ is an $L \times n$ matrix where $Q_{ji} = q_i^j \pmod{p_i}$.
    \item $I$ is an $L \times L$ identity matrix representing the coefficients of $d$.
    \item $P$ is an $n \times n$ diagonal matrix where $P_{ii} = p_i$, representing the moduli constraints.
    \item $0$ is an $n \times L$ zero matrix.
\end{itemize}

\subsection{Reduction and Reconstruction}
The rows of $M$ span a lattice. A vector $v$ in this lattice has the form:
\[
v = (\lambda \cdot R, d)
\]
where $R$ represents the remainder of the polynomial evaluation modulo $p_i$. We aim to find a vector where $R=0$ (meaning the hash difference is a multiple of the modulus) and $d$ is small (non-zero but within character range).

By multiplying the top-left and bottom-left blocks by a sufficiently large $\lambda$, we penalize any non-zero remainder $R$. We then apply the \textbf{$L^2$ reduction algorithm} (a variant of LLL) to reduce the basis $M$. The shortest vector in the reduced basis typically yields a vector with $R=0$ and a valid difference array $d$. From $d$, we construct strings $a$ and $b$ by setting $a_j = \max(0, d_j)$ and $b_j = \max(0, -d_j)$ (shifted by a base character), guaranteeing a collision.

\subsection{Birthday Attack Strategy}
\label{appx:birthday_paradox}

While Lattice Reduction is effective for polynomial rolling hashes with large moduli (e.g., $2^{64}$), it requires the hash function to have a specific linear structure. For generic hash functions or smaller moduli (e.g., $M \approx 10^9$), we utilize a probabilistic approach based on the \textbf{Birthday Paradox}.

The Birthday Paradox states that in a set of $n$ randomly chosen elements, where each element is drawn uniformly from a domain of size $M$, the probability that at least two elements are identical (a collision) is approximately:
\[
P(\text{collision}) \approx 1 - e^{-\frac{n(n-1)}{2M}} \approx 1 - e^{-\frac{n^2}{2M}}
\]
To achieve a collision probability of $P \approx 0.5$ (50\%), the required number of generated test cases $n$ is:
\[
n \approx \sqrt{2M \ln 2} \approx 1.177 \sqrt{M}
\]
For a typical 32-bit integer hash ($M \approx 4 \times 10^9$), the agent only needs to generate approximately $n \approx 75,000$ inputs to find a collision with high confidence. This is computationally trivial for the CodeHacker agent. 

\paragraph{Implementation:}
The agent generates two sets of random strings, $S_A$ and $S_B$. It computes the hash values for all strings in $S_A$ and stores them in a hash map. Then, it computes hashes for strings in $S_B$ and checks for existence in the map. This \textit{meet-in-the-middle} approach efficiently discovers collisions for single-hash checks commonly found in weaker solutions.

\section{Case Study}

\subsection{A Weak Checker}
\label{appx:weak_checker}

In this section, we analyze a weak checker from the problem \textit{Codeforces 25\_B}. The problem allows multiple valid output formats, making strict string comparison insufficient.

\subsubsection{Problem: Phone Numbers}
\textbf{Description:} Given a string of $n$ digits ($2 \le n \le 100$), divide it into groups of length 2 or 3, separated by hyphens ('-').
\newline
\textbf{Output Requirements:}
\begin{itemize}
    \item The output must contain only digits and hyphens.
    \item Every group must have a length of exactly 2 or 3.
    \item The concatenation of the groups (removing hyphens) must exactly match the original string.
\end{itemize}

\subsubsection{Vulnerability Analysis}
The original weak checker (Figure~\ref{fig:weak-checker}) attempts to validate the output by splitting the string by hyphens. However, it lacks robust parsing logic for edge cases:
\begin{enumerate}
    \item It may crash or misbehave if the output contains consecutive hyphens or leading/trailing hyphens.
    \item It fails to rigorously check for non-digit characters (e.g., whitespace or letters) in some implementations.
\end{enumerate}
Our refined checker (Figure~\ref{fig:our-checker}) performs character-level validation, ensuring that no illegal characters exist and that the structure strictly follows the grouping rules before content verification.

\begin{figure*}[t]
\centering
\begin{minipage}{0.95\textwidth}
\begin{lstlisting}
#include "testlib.h"
#include <bits/stdc++.h>
using namespace std;
void readAndCheckAnswer(InStream& stream, const string& pn, const string& who) {
    string line = stream.readLine();
    // Remove spaces
    line.erase(remove(line.begin(), line.end(), ' '), line.end());
    // Split line into groups on '-'
    vector<string> groups;
    stringstream ss(line);
    string group;
    while (getline(ss, group, '-')) {
        groups.push_back(group);
    }
    // For each group, check that length is 2 or 3 and consists of digits
    for (size_t i = 0; i < groups.size(); ++i) {
        if (groups[i].length() < 2 || groups[i].length() > 3) {
            stream.quitf(_wa, "%s: group %d has invalid length %zu (should be 2 or 3)", who.c_str(), int(i + 1), groups[i].length());
        }
        if (!all_of(groups[i].begin(), groups[i].end(), ::isdigit)) {
            stream.quitf(_wa, "%s: group %d contains non-digit characters", who.c_str(), int(i + 1));
        }
    }
    // Concatenate all groups and check if it matches pn
    string reconstructed;
    for (const auto& g : groups) {
        reconstructed += g;
    }
    if (reconstructed != pn) {
        stream.quitf(_wa, "%s: concatenated groups do not match original phone number", who.c_str());
    }
}
int main(int argc, char* argv[]) {
    registerTestlibCmd(argc, argv);
    int n = inf.readInt(2, 100);
    string pn = inf.readToken(format("[0-9]{%d}", n), "phone number");
    readAndCheckAnswer(ans, pn, "jury's answer");
    readAndCheckAnswer(ouf, pn, "participant's answer");
    quitf(_ok, "Correct answer");
}

\end{lstlisting}
\end{minipage}
\caption{Weak Checker in $CodeContest^+$ for problem Codeforces 25\_B}
\label{fig:weak-checker}
\end{figure*}

\begin{figure*}[t]
\centering
\begin{minipage}{0.95\textwidth}
\begin{lstlisting}
#include "testlib.h"
#include <bits/stdc++.h>
using namespace std;
void readAndCheckAnswer(InStream& stream, const string& pn, const string& who) {
    string line = stream.readLine();
    if (line.empty()) {
        stream.quitf(_wa, "%s: empty output", who.c_str());
    }
    // Validate characters: no whitespace allowed, only digits and '-'
    for (size_t i = 0; i < line.size(); ++i) {
        unsigned char uc = static_cast<unsigned char>(line[i]);
        if (std::isspace(uc)) {
            stream.quitf(_wa, "%s: output contains whitespace characters", who.c_str());
        }
        if (line[i] != '-' && !std::isdigit(uc)) {
            // Print the offending character; if non-printable, show its code
            if (std::isprint(uc)) {
                stream.quitf(_wa, "%s: output contains invalid character '%c'", who.c_str(), line[i]);
            } else {
                stream.quitf(_wa, "%s: output contains invalid character with code %d", who.c_str(), int(uc));
            }
        }
    }
    // Split line into groups on '-'
    vector<string> groups;
    string cur;
    for (char c : line) {
        if (c == '-') {
            groups.push_back(cur);
            cur.clear();
        } else {
            cur.push_back(c);
        }
    }
    groups.push_back(cur);
    // For each group, check that length is 2 or 3 and consists of digits
    for (size_t i = 0; i < groups.size(); ++i) {
        int len = int(groups[i].length());
        if (len < 2 || len > 3) {
            stream.quitf(_wa, "%s: group %d has invalid length %d (should be 2 or 3)", who.c_str(), int(i + 1), len);
        }
        if (!all_of(groups[i].begin(), groups[i].end(), [](char ch) { return std::isdigit(static_cast<unsigned char>(ch)); })) {
            stream.quitf(_wa, "%s: group %d contains non-digit characters", who.c_str(), int(i + 1));
        }
    }
    // Concatenate all groups and check if it matches pn
    string reconstructed;
    for (const auto& g : groups) {
        reconstructed += g;
    }
    if (reconstructed != pn) {
        stream.quitf(_wa, "%s: concatenated groups do not match original phone number", who.c_str());
    }
}
int main(int argc, char* argv[]) {
    registerTestlibCmd(argc, argv);
    int n = inf.readInt(2, 100);
    string pn = inf.readToken(format("[0-9]{%d}", n), "phone number");
    readAndCheckAnswer(ans, pn, "jury's answer");
    readAndCheckAnswer(ouf, pn, "participant's answer");
    quitf(_ok, "Correct");
}

\end{lstlisting}
\end{minipage}
\caption{Our Checker for problem Codeforces 25\_B}
\label{fig:our-checker}
\end{figure*}

\subsection{A Wrong Validator}

In this section, we examine a validator flaw in \textit{Codeforces 309\_C}. The validator restricts input values more strictly than the problem statement allows, causing valid test cases to be rejected.

\subsubsection{Problem: Memory Management}
\textbf{Description:} You are given an array $A$ of $N$ integers and an array $B$ of $M$ integers. Each element $b_j \in B$ represents a memory block of size $2^{b_j}$.
\newline
\textbf{Constraints:}
\begin{itemize}
    \item $1 \le N, M \le 10^6$.
    \item Elements of $A$: $1 \le a_i \le 10^9$.
    \item Elements of $B$: $0 \le b_j \le 29$. (Crucially, $2^0=1$ is valid).
\end{itemize}

\subsubsection{Vulnerability Analysis}
The original validator (Figure~\ref{fig:wrong-validator}) incorrectly enforces the range $[1, 60]$ for array $B$. This logic explicitly forbids $b_j=0$, effectively disallowing memory blocks of size $2^0=1$. However, the problem statement allows $b_j=0$.
This False Negative error prevents the system from testing edge cases involving the smallest unit of memory, potentially masking bugs in solutions that fail to handle size 1 blocks.

Figure~\ref{fig:wrong-validator} shows the original validator code in CodeContest\pl. In this validator, the program checks the value of each element in $b$ to ensure that $2^{b_j} \leq 10^9$. However, the range for $b_j$ is incorrectly limited to $[1, 60]$, excluding the case where $b_j = 0$. This results in the rejection of valid test cases where $b_j = 0$.

\begin{figure*}[t]
\centering
\begin{minipage}{0.95\textwidth}
\begin{lstlisting}
#include "testlib.h"
#include <bits/stdc++.h>
using namespace std;
int main(int argc, char* argv[]) {
    registerValidation(argc, argv);
    int n = inf.readInt(1, 1000000, "n");
    inf.readSpace();
    int m = inf.readInt(1, 1000000, "m");
    inf.readEoln();
    vector<int> a = inf.readInts(n, 1, 1000000000, "a_i");
    inf.readEoln();
    vector<int> b = inf.readInts(m, 1, 60, "b_j");
    inf.readEoln();
    for (int i = 0; i < m; i++) {
        long long s = 1LL << b[i];
        ensuref(s <= 1000000000LL, "2^%d is %lld, which is greater than 1e9", b[i], s);
    }
    inf.readEof();
    return 0;
}
\end{lstlisting}
\end{minipage}
\caption{Wrong Validator in CodeContest\pl for problem Codeforces 309\_C}
\label{fig:wrong-validator}
\end{figure*}

In our revised version of the validator, shown in Figure~\ref{fig:correct-validator}, we have updated the validator to allow for $b_j = 0$ by adjusting the input range for $b_j$ from $[0, 60]$ to $[0, 29]$. This change ensures that the case where $b_j = 0$ is now accepted, thus allowing $2^0 = 1$ to pass the validation. This adjustment corrects the original validator's oversight and ensures that all valid test cases are processed correctly.

\begin{figure*}[t]
\centering
\begin{minipage}{0.95\textwidth}
\begin{lstlisting}
#include "testlib.h"
#include <bits/stdc++.h>
using namespace std;
int main(int argc, char* argv[]) {
    registerValidation(argc, argv);
    int n = inf.readInt(1, 1000000, "n");
    inf.readSpace();
    int m = inf.readInt(1, 1000000, "m");
    inf.readEoln();
    vector<int> a = inf.readInts(n, 1, 1000000000, "a_i");
    inf.readEoln();
    vector<int> b = inf.readInts(m, 0, 29, "b_j");
    inf.readEoln();
    inf.readEof();
    return 0;
}
\end{lstlisting}
\end{minipage}
\caption{Our Corrected Validator for problem Codeforces 309\_C}
\label{fig:correct-validator}
\end{figure*}

This case study illustrates the importance of properly defining the valid input ranges and corner cases in the validator. The original error highlights how minor oversights in constraint handling can lead to unnecessary test failures, affecting the overall reliability of the contest system. By fixing this, we ensure that all valid edge cases are properly accepted, making the validator more robust and reliable.

\subsection{Another Wrong Validator}
This case study uses \textit{Codeforces 177\_C2} to illustrate a discrepancy where the validator is too permissive compared to the problem statement, potentially leading to Time Limit Exceeded (TLE) on valid solutions.

\subsubsection{Problem: Party}
\textbf{Description:} A social network with $N$ people. There are $k$ pairs of friends and $m$ pairs of enemies. Find the maximum size of a valid party group.
\newline
\textbf{Constraints:}
\begin{itemize}
    \item $N \le 2000$.
    \item $k \le 100,000$ (The number of friendship pairs).
    \item The graph of friends must be simple (no self-loops, no duplicate edges).
\end{itemize}

\subsubsection{Vulnerability Analysis}
The problem statement explicitly limits $k$ to $100,000$. However, the maximum possible edges in a graph of $N=2000$ nodes is $N(N-1)/2 \approx 2 \times 10^6$.
The original validator (Figure~\ref{fig:wrong-validator2}) calculates the theoretical maximum edges (\texttt{max\_edges}) and allows $k$ to go up to this value.
This means the validator accepts test cases with up to 2 million edges, far exceeding the stated limit of 100,000. Solutions optimized for $k=10^5$ (e.g., using adjacency lists) might unexpectedly TLE when fed $2 \times 10^6$ edges.

\begin{figure*}[t]
\centering
\begin{minipage}{0.95\textwidth}
\begin{lstlisting}
#include "testlib.h"
#include <bits/stdc++.h>
using namespace std;
int main(int argc, char* argv[]) {
    registerValidation(argc, argv);
    int n = inf.readInt(2, 2000, "n");
    inf.readEoln();
    int64_t max_edges = int64_t(n) * (n - 1) / 2;
    int k = inf.readInt(0, max_edges, "k");
    inf.readEoln();
    set<pair<int,int>> friendship_pairs;
    for (int i = 0; i < k; i++) {
        int u = inf.readInt(1, n, "u_i");
        inf.readSpace();
        int v = inf.readInt(1, n, "v_i");
        inf.readEoln();
        ensuref(u != v, "A person cannot be friends with themselves: u=%d, v=%d", u, v);
        int x = min(u, v);
        int y = max(u, v);
        pair<int,int> p = make_pair(x, y);
        ensuref(friendship_pairs.count(p) == 0, "Duplicate friendship pair (%d, %d)", x, y);
        friendship_pairs.insert(p);
    }
    int m = inf.readInt(0, max_edges, "m");
    inf.readEoln();
    set<pair<int,int>> dislike_pairs;
    for (int i = 0; i < m; i++) {
        int u = inf.readInt(1, n, "u_i");
        inf.readSpace();
        int v = inf.readInt(1, n, "v_i");
        inf.readEoln();
        ensuref(u != v, "A person cannot dislike themselves: u=%d, v=%d", u, v);
        int x = min(u, v);
        int y = max(u, v);
        pair<int,int> p = make_pair(x, y);
        ensuref(dislike_pairs.count(p) == 0, "Duplicate dislike pair (%d, %d)", x, y);
        ensuref(friendship_pairs.count(p) == 0, "Pair (%d, %d) cannot be both friends and dislike each other", x, y);
        dislike_pairs.insert(p);
    }
    inf.readEof();
    return 0;
}
\end{lstlisting}
\end{minipage}
\caption{Wrong Validator in CodeContest\pl for problem Codeforces 177\_{C2}}
\label{fig:wrong-validator2}
\end{figure*}

In our revised version of the validator, shown in Figure~\ref{fig:correct-validator2}, we have updated the validator to properly handle the $k$ constraint. Specifically, we now enforce that $k$ cannot exceed $100000$, as stated in the problem description. This ensures that the validator only accepts valid test cases and prevents the Time Limit Exceeded (TLE) errors caused by excessively large values of $k$.

\begin{figure*}[t]
\centering
\begin{minipage}{0.95\textwidth}
\begin{lstlisting}
#include "testlib.h"
#include <bits/stdc++.h>
using namespace std;
int main(int argc, char* argv[]) {
    registerValidation(argc, argv);
    int n = inf.readInt(2, 2000, "n");
    inf.readEoln();
    int64_t max_edges = min(100000ll, int64_t(n) * (n - 1) / 2);
    int k = inf.readInt(0, max_edges, "k");
    inf.readEoln();
    set<pair<int,int>> friendship_pairs;
    for (int i = 0; i < k; i++) {
        int u = inf.readInt(1, n, "u_i");
        inf.readSpace();
        int v = inf.readInt(1, n, "v_i");
        inf.readEoln();
        ensuref(u != v, "A person cannot be friends with themselves: u=%d, v=%d", u, v);
        int x = min(u, v);
        int y = max(u, v);
        pair<int,int> p = make_pair(x, y);
        ensuref(friendship_pairs.count(p) == 0, "Duplicate friendship pair (%d, %d)", x, y);
        friendship_pairs.insert(p);
    }
    int m = inf.readInt(0, max_edges, "m");
    inf.readEoln();
    set<pair<int,int>> dislike_pairs;
    for (int i = 0; i < m; i++) {
        int u = inf.readInt(1, n, "u_i");
        inf.readSpace();
        int v = inf.readInt(1, n, "v_i");
        inf.readEoln();
        ensuref(u != v, "A person cannot dislike themselves: u=%d, v=%d", u, v);
        int x = min(u, v);
        int y = max(u, v);
        pair<int,int> p = make_pair(x, y);
        ensuref(dislike_pairs.count(p) == 0, "Duplicate dislike pair (%d, %d)", x, y);
        ensuref(friendship_pairs.count(p) == 0, "Pair (%d, %d) cannot be both friends and dislike each other", x, y);
        dislike_pairs.insert(p);
    }
    inf.readEof();
    return 0;
}
\end{lstlisting}
\end{minipage}
\caption{Our Corrected Validator for problem Codeforces 177\_{C2}}
\label{fig:correct-validator2}
\end{figure*}

This case study highlights the importance of aligning the validator with the problem's constraints to avoid unnecessary errors. By ensuring that $k$ does not exceed the value of 100,000, as specified in the problem description, we ensure that all valid test cases are accepted while preventing performance issues and TLE errors. This adjustment improves the reliability of the validator and ensures consistency with the problem's defined limits.

\clearpage
\newpage
\section{Case Study: Heuristic Failure in Number Theory}
\label{appx:quadratic_set}

In this section, we examine a case where a solution relies on flawed mathematical heuristics and floating-point arithmetic for a number theory problem. This case study illustrates how \textbf{CodeHacker} can detect vulnerabilities that arise from false generalizations, which are often missed by weak random test cases.

\subsection{Problem: Quadratic Set}
The problem asks us to find the maximum size subset of $\{1, 2, \dots, n\}$ such that the product of their factorials is a perfect square. The constraints are $n \le 10^6$.

A correct solution typically requires examining the prime factorization of the factorials (specifically the parity of prime exponents) using techniques like XOR hashing (Zobrist Hashing). However, the submission below attempts to solve the problem using a greedy heuristic based on $n \pmod 4$ and simple floating-point square checks.

\subsection{The Vulnerable Submission}
The code in Figure~\ref{fig:heuristic-fail} attempts to guess which elements to remove (at most 2) to make the product a square. It relies on the helper function \texttt{is\_perfect\_square} to check conditions derived from loose patterns observed in small numbers.

\begin{figure*}[h]
\centering
\begin{minipage}{0.95\textwidth}
\begin{lstlisting}[style=cppstyle]
#include <bits/stdc++.h>
using namespace std;

// Flaw 1: Floating point precision issues and logic error
// This function checks if a number is a perfect square using sqrt
// which is unreliable for large integers and irrelevant for the
// factorial product property required by the problem.
bool is_perfect_square(unsigned long long int n) {
  if (pow((long int)(sqrt(n)), 2) == n) {
    return true;
  }
  return false;
}

void solve() {
  unsigned long int n;
  cin >> n;
  vector<unsigned long int> arr = {};
  unsigned long int halfi = (long int)n / 2;
  
  // Flaw 2: Heuristic logic based on modulo 4
  // The code assumes that the set of removed numbers can always
  // be found among {halfi, n, halfi+1, 2, ...} based on n % 4.
  // This generalization holds for small N but fails for complex cases.
  unsigned long long int asdf = (unsigned long long int)2 * (halfi) * (halfi - 1);
  
  if (remainderf(n, 4) == 0) {
    arr = {halfi};
  } else if (n == 1) {
    arr = {};
  } else if (remainderf(n, 4) == 1) {
    arr = {halfi, n};
  } else if ((n % 4) == 2) {
    if (is_perfect_square((unsigned long long int)(n + 2)))
      arr = {halfi + 1};
    else if (is_perfect_square((unsigned long long int)(n * (halfi - 1))))
      arr = {halfi - 2};
    else
      arr = {halfi, 2};
  } else {
    // This branch handles n % 4 == 3
    if (is_perfect_square((unsigned long long int)(n + 1)))
      arr = {halfi + 1, n};
    else if (is_perfect_square(asdf))
      arr = {n, halfi - 2};
    else if (is_perfect_square((halfi - 1) * n))
      arr = {halfi - 2, n - 2};
    else
      arr = {2, halfi, n};
  }
  
  vector<int> ans = {};
  for (int i = 1; i < n + 1; i++) {
    if (find(arr.begin(), arr.end(), i) == arr.end()) ans.push_back(i);
  }
  cout << ans.size() << endl;
  for (int el = 0; el < ans.size(); el++) {
    cout << ans[el] << ' ';
  }
  cout << endl;
}

int main() {
  solve();
  return 0;
}
\end{lstlisting}
\end{minipage}
\caption{Submission for "Quadratic Set" using flawed heuristics.}
\label{fig:heuristic-fail}
\end{figure*}

\subsection{Vulnerability Analysis}
This submission exhibits a \textbf{Logic Error} rooted in a \textbf{False Generalization}.
\begin{itemize}
    \item \textbf{Mathematical Flaw:} The condition $\prod a_i! = k^2$ depends on the parity of prime factors. The code tries to satisfy this by checking if specific arithmetic combinations of $N$ (e.g., $N+2$ or $N(N/2-1)$) are perfect squares. There is no number-theoretic basis guaranteeing that these specific checks cover all cases where the factorial product becomes a square.
    \item \textbf{Floating Point Risk:} The use of \texttt{remainderf} and \texttt{pow/sqrt} introduces precision risks, although the primary failure here is logical.
\end{itemize}

\paragraph{The Hack.}
Standard small test cases often satisfy the modulo patterns hardcoded in the solution. However, CodeHacker's \textbf{Stress Test} module, configured to explore high-magnitude inputs with specific prime properties, identifies $N = 998787$ as a failure case.
For $N = 998787$, the code enters the \texttt{else} branch ($998787 \pmod 4 = 3$). The heuristics fail to find the optimal removal set, or incorrecty identify the removal set due to the lack of rigor in the `is\_perfect\_square` logic applied to non-factorial numbers. The correct solution requires a hashing approach to track the parity of prime factors up to $N$, which this code completely lacks.

\clearpage
\newpage
\section{Case Study: Logical Oversight in Construction}
\label{appx:logic_hack}

In this section, we present a successful hack generated by CodeHacker against a heuristic solution for \textit{Codeforces 1388A}. This case illustrates how the agent identifies corner cases where a greedy construction strategy violates distinctness constraints.

\subsection{Problem: Captain Flint and Crew Recruitment}
\textbf{Description:} Given an integer $N$, represent it as the sum of 4 \textbf{distinct} positive integers, such that at least 3 of them are "nearly prime".
\newline
\textbf{Definitions:}
\begin{itemize}
    \item A number is "nearly prime" if it is the product of two distinct prime numbers (e.g., $6=2\cdot3$, $10=2\cdot5$, $14=2\cdot7$).
\end{itemize}

\subsection{Vulnerability Analysis}
The submitted code (Figure~\ref{fig:logic-fail}) adopts a static greedy strategy. It always attempts to use the three smallest nearly primes: 6, 10, and 14. Their sum is $30$.
The code logic sets the fourth number to be $x = N - 30$.
\newline
\textbf{The Flaw:} The problem requires all 4 integers to be \textbf{different}. The code fails to check if the calculated fourth number $x$ coincides with one of the fixed numbers $\{6, 10, 14\}$.
\begin{itemize}
    \item If $x = 6 \implies N = 36$, output is \texttt{6 10 14 6} (Duplicate 6).
    \item If $x = 10 \implies N = 40$, output is \texttt{6 10 14 10} (Duplicate 10).
    \item If $x = 14 \implies N = 44$, output is \texttt{6 10 14 14} (Duplicate 14).
\end{itemize}
CodeHacker successfully identified these collision points and generated the adversarial inputs $N \in \{36, 40, 44\}$.

\begin{figure*}[h]
\centering
\begin{minipage}{0.95\textwidth}
\begin{lstlisting}[style=cppstyle]
#include <bits/stdc++.h>
using namespace std;

int main() {
    int t; cin >> t;
    while (t--) {
        int n; cin >> n;
        // The code correctly identifies that any N <= 30 is impossible
        // because 6+10+14+1 = 31 is the minimum possible sum.
        if (n < 31) {
            cout << "NO\n";
        } else {
            cout << "YES\n";
            // BUG: This construction assumes the 4th number (n-30)
            // will never collide with 6, 10, or 14.
            // For N=36, it prints "6 10 14 6", which has duplicates.
            cout << "6 10 14 " << n - 30 << "\n";
        }
    }
}
\end{lstlisting}
\end{minipage}
\caption{The vulnerable submission fails to handle collision cases.}
\label{fig:logic-fail}
\end{figure*}

\paragraph{Correct Approach:}
A robust solution must handle these collisions. For example, if $N-30$ causes a collision (e.g., $N=36$), one can replace the fixed set $\{6, 10, 14\}$ (sum 30) with $\{6, 10, 15\}$ (sum 31), making the fourth number $N-31$. For $N=36$, this yields \texttt{6 10 15 5}, which are all distinct.

\newpage
\onecolumn
\section{Prompt Templates\label{appx:prompts}}

In this section, we present some prompt templates.

\begin{promptbox}[Checker Hack Generator Prompt]
\small
Please review the provided checker code and identify logical flaws. Each flaw should target a different logical issue that causes the checker to incorrectly judge the output.

\vspace{0.5em}
\noindent\textbf{\# Inputs} \\
\textbf{Problem Description:} \texttt{\{problem\_description\}} \\
\textbf{Checker Code:} \texttt{\{checker\_code\}}

\vspace{0.5em}
\noindent\textbf{\# Reasoning Steps}
\begin{enumerate}
    \setlength\itemsep{0em}
    \item \textbf{Understand Requirements}: Analyze the problem to determine if multiple valid solutions are allowed (Special Judge).
    \item \textbf{Analyze Checker Logic}: Look for missing checks (too permissive) or overly rigid comparisons against standard output (too strict).
    \item \textbf{Construct Hack}:
    \begin{itemize}
        \item \textbf{False Positive}: Construct an \textit{invalid} output that the checker accepts.
        \item \textbf{False Negative}: Construct a \textit{valid} alternative output (different from std) that the checker rejects.
    \end{itemize}
    \item \textbf{Verify}: Ensure your constructed output strictly follows the format.
\end{enumerate}

\vspace{0.5em}
\noindent\textbf{\# Target Bug Types}
\begin{description}
    \setlength\itemsep{0em}
    \item[Type 1: False Positive (Too Permissive)] \hfill \\
    The checker wrongly \textbf{ACCEPTS} an invalid output.
    \item[Type 2: False Negative (Too Strict)] \hfill \\
    The checker wrongly \textbf{REJECTS} a valid output. This often happens when the checker requires the output to be identical to the standard solution, ignoring other valid permutations or solutions.
\end{description}

\vspace{0.5em}
\noindent\textbf{\# Output Format (JSON)}
\begin{lstlisting}[style=jsonstyle]
{
  "test_cases": [
    {
      "strategy": "Checker ignores array sorting (False Positive)",
      "bug_type": "false_positive",
      "test_input": "<INPUT>",
      "fake_output": "<WRONG_OUTPUT_ACCEPTED>"
    },
    {
      "strategy": "Checker rejects valid reverse order (False Negative)",
      "bug_type": "false_negative",
      "test_input": "<INPUT>",
      "fake_output": "<VALID_ALT_OUTPUT_REJECTED>"
    }
  ]
}
\end{lstlisting}
\end{promptbox}

\begin{promptbox}[Validator Hack Generator Prompt]
\small
Please review the provided validator code and identify logical flaws. Each flaw should target a different logical issue that causes the validator to incorrectly judge the input case.

\vspace{0.5em}
\noindent\textbf{\# Inputs} \\
\textbf{Constraints Description:} \texttt{\{problem\_description\}} (Focus on Input Format and Data Ranges) \\
\textbf{Validator Code:} \texttt{\{validator\_code\}}

\vspace{0.5em}
\noindent\textbf{\# Reasoning Steps}
\begin{enumerate}
    \setlength\itemsep{0em}
    \item \textbf{Extract Constraints}: List every constraint from the text (e.g., $1 \le N \le 10^5$, graph is connected, no duplicate edges).
    \item \textbf{Audit Code}: Check if the Validator enforces each constraint using \texttt{ensuref()} or strict reading methods.
    \item \textbf{Construct Hack}:
    \begin{itemize}
        \item \textbf{Type 1 (False Positive)}: Generate an \textbf{ILLEGAL} input that the Validator likely \textbf{ACCEPTS} (e.g., Validator uses \texttt{int} but input is $10^{10}$).
        \item \textbf{Type 2 (False Negative)}: Generate a \textbf{LEGAL} input that the Validator likely \textbf{REJECTS} (e.g., Validator checks $N < 100$ but text says $N \le 100$).
    \end{itemize}
\end{enumerate}

\vspace{0.5em}
\noindent\textbf{\# Bug Types}
\begin{description}
    \setlength\itemsep{0em}
    \item[False Positive (Too Loose)] The Validator accepts an input that violates the constraints. (This is dangerous as it crashes solutions).
    \item[False Negative (Too Strict)] The Validator rejects a perfectly valid input.
\end{description}

\noindent\textbf{\# Output Format (JSON)}
\begin{lstlisting}[style=jsonstyle]
{
  "test_cases": [
    {
      "strategy": "The problem requires a DAG (no cycles), but the validator only checks connectivity. This input contains a cycle.",
      "bug_type": "false_positive", 
      "test_input": "3 3\n1 2\n2 3\n3 1",
      "expected_validity": "invalid" // This input SHOULD be invalid
    }
  ]
}
\end{lstlisting}
\end{promptbox}

\begin{promptbox}[Code Analyst System Prompt]
\small
You are an expert Code Auditor for Competitive Programming. Your goal is to identify bugs (WA, TLE, RE, MLE) in a C++ submission by interacting with an execution environment.

\vspace{0.5em}
\noindent\textbf{\# Communication Protocol}
This is a multi-turn dialogue. In each turn, you must output a **Thought** followed by a **Tool Call** (or a Final Report).
The system will execute your tool and return the \texttt{[Tool Output]} in the next message.

\vspace{0.5em}
\noindent\textbf{\# Available Tools}
\begin{description}
    \setlength{\itemsep}{0em}
    \item[\texttt{run\_python(script\_code)}] \mbox{} \\
    Executes \textbf{arbitrary Python code}. You can use loops (\texttt{for/while}), define functions, imports (math, itertools), etc.
    \item[\texttt{run\_cpp(input\_content)}] \mbox{} \\
    Compiles and runs the target C++ code with the provided \texttt{input\_content} (string).
    Returns: \texttt{stdout}, \texttt{stderr}, and \texttt{ExitCode}.
    \item[\texttt{finish(code\_analysis\_report)}] \mbox{} \\
    Submits the final vulnerability report.
\end{description}

\vspace{0.5em}
\noindent\textbf{\# One-Shot Example}
\begin{quote}
\textbf{User:} Problem: Sum of squares $1^2 + \dots + N^2$. $N \le 10^6$. Code uses \texttt{int}.
\textbf{Assistant:}
Thought: The formula is $N(N+1)(2N+1)/6$. I need to check if $N=10^6$ overflows a 32-bit signed int ($2 \times 10^9$).

\textbf{Call:}
\begin{lstlisting}[language=Python, numbers=none, frame=none, basicstyle=\ttfamily\small, xleftmargin=1em]
run_python("""
sum = 0
for i in range(1, 10000001):
    sum += i * i
print(sum)
""")
\end{lstlisting}

\textbf{User:} [Tool Output] 333333833333500000

\textbf{Assistant:}
Thought: The result is $\approx 3.3 \times 10^{17}$, which far exceeds $2 \times 10^9$. This confirms Overflow. Now I'll probe the code with a small case to see if logic is correct otherwise. Expected: $1^2+2^2+3^2 = 14$.

Call: \texttt{run\_cpp("3")}

\textbf{User:} [Tool Output] Stdout: "14", ExitCode: 0

\textbf{Assistant:}
Thought: Small case works, but large case will definitely overflow.

Call: \texttt{finish("The code implements the summation formula correctly but risks overflow for $N$ close to upper bounds. While $N=10^6$ fits, any intermediate calculation without modulo might exceed limits. The vulnerability is Integer Overflow on large inputs.")}
\end{quote}

\vspace{0.5em}
\noindent\textbf{\# Your Task} \\
\textbf{Problem:} \texttt{\{problem\_description\}} \\
\textbf{Code:} \texttt{\{target\_code\}}
\end{promptbox}

\begin{promptbox}[Stress Test Generator Prompt]
\small
You are an expert in Competitive Programming. Your task is to write a **Randomized Test Case Generator** in C++.
This generator will be used for "Stress Testing" (Fuzzing) a target solution.

\vspace{0.5em}
\noindent\textbf{\# Problem Description} \\
\texttt{\{problem\_description\}}

\vspace{0.5em}
\noindent\textbf{\# Your Task} \\
Write a complete C++ program that prints a SINGLE valid test case to standard output.

\noindent\textbf{\# Requirements for the Generator Code:}
\begin{enumerate}
    \setlength\itemsep{0em}
    \item \textbf{Randomization}: Initialize the random seed using \texttt{mt19937 rng(chrono::steady\_clock::now().time\_since\_epoch().count());} or \texttt{srand(time(0))}.
    \item \textbf{Validity}: The output MUST strictly follow the problem's input format and constraints (e.g., if the graph must be a tree, ensure no cycles and connectivity).
    \item \textbf{Scale}: Bias the generation towards \textbf{large inputs} (close to the maximum $N$) to test for Time Limit Exceeded (TLE) errors, but occasionally generate small edge cases.
    \item \textbf{Robustness}: Avoid undefined behavior in your generator (e.g., ensure \texttt{rand() \% 0} never happens).
\end{enumerate}

\vspace{0.5em}
\noindent\textbf{\# Helper Functions Recommendation} \\
You may define helper functions like \texttt{long long rand(long long a, long long b)} to generate numbers in range $[a, b]$.

\vspace{0.5em}
\noindent\textbf{\# Output Format} \\
Provide ONLY the C++ code block. Start with \texttt{\#include <bits/stdc++.h>}.
\end{promptbox}

\begin{promptbox}[Generator Prompt: Bug Discovery \& Test Case Generation]
\small 

You are an expert in competitive programming. Your task is to find a bug in the following code and generate a test case that will expose it.

\vspace{0.5em}
\noindent\textbf{\# Problem Description:} \\
\texttt{\{problem\_description\}}

\vspace{0.5em}
\noindent\textbf{\# Incorrect Code:}
\begin{lstlisting}[style=cppstyle]
{incorrect_code}
\end{lstlisting}

\vspace{0.5em}
\noindent\textbf{\# Hash Collision Data (Optional):} \\
Our anti-hash generator has found two strings which have same hash values.
\texttt{\{Hash Collision Data\}}

\vspace{0.5em}
\noindent\textbf{\#\#\# Your Task}

\textbf{Construct a test case (input) that:}
\begin{enumerate}
    \setlength\itemsep{0em}
    \item Satisfies all the constraints of the original problem
    \item Will cause the buggy implementation to produce an \textbf{incorrect result}
    \item A correct solution would handle properly
\end{enumerate}

\noindent\textbf{\#\#\# Types of Errors to Target}

The buggy code may fail in one of these ways:
\begin{itemize}
    \setlength\itemsep{0em}
    \item \textbf{WA (Wrong Answer)}: Produces incorrect output due to logical errors, wrong algorithms, or edge case mishandling
    \item \textbf{TLE (Time Limit Exceeded)}: Takes too long to execute, often due to inefficient algorithms or infinite loops
    \item \textbf{MLE (Memory Limit Exceeded)}: Requires too much memory space, often due to increasing space needed in the program.
    \item \textbf{RE (Runtime Error)}: Crashes during execution due to array out of bounds, division by zero, stack overflow, etc.
\end{itemize}
Your test case should be designed to trigger one of these error types.

\vspace{0.5em}
\noindent\textbf{\# Output Format:} \\
Provide a complete C++ program that generates the test input. The program should:
\begin{itemize}
    \setlength\itemsep{0em}
    \item Always include \texttt{bits/stdc++.h} first
    \item Have a \texttt{main()} function that outputs the test case
    \item Follow the input format specified in the problem
\end{itemize}

\noindent\textbf{\#\#\# Example Output Format}

Your generator can be as simple or complex as needed. Here are examples:

\textbf{Simple Example (Direct Output):}
\begin{lstlisting}[style=cppstyle]
#include <bits/stdc++.h>
using namespace std;

int main() {
    // Bug: code assumes n <= 1000, but constraint allows up to 10^5
    // This test case uses n = 100000 to trigger array overflow
    cout << "100000" << endl;
    return 0;
}
\end{lstlisting}

\textbf{Complex Example (Algorithmic Generation):}
\begin{lstlisting}[style=cppstyle]
#include <bits/stdc++.h>
using namespace std;

int main() {
    // Bug: O(n^2) algorithm will TLE on large inputs
    // Generate maximum size test case with worst-case pattern
    int n = 200000;
    cout << n << endl;

    // Generate adversarial pattern: descending order
    for (int i = n; i >= 1; i--) {
        cout << i;
        if (i > 1) cout << " ";
    }
    cout << endl;
    return 0;
}
\end{lstlisting}

\textbf{Note:} Your generator can include loops, conditionals, calculations, or any logic needed to construct the test case.

\vspace{0.5em}
\noindent\textbf{\#\#\# Additional Guidelines}
\begin{enumerate}
    \setlength\itemsep{0em}
    \item \textbf{Analyze the bug}: First identify what type of bug exists (Logic errors $\to$ WA, Inefficient loops $\to$ TLE, etc.)
    \item \textbf{Target the weakness}: Design your test case to specifically trigger that bug
    \item \textbf{Stay within constraints}: Your test case must be valid according to the problem constraints
    \item \textbf{Maximize impact}: Choose values that make the error obvious and deterministic
\end{enumerate}

\vspace{0.5em}
\noindent\textbf{Now, write the C++ generator based on this advice. Think step by step.}

\end{promptbox}

\begin{promptbox}[Anti-Hash Generator]
\small
You are a Cryptanalysis and Competitive Programming Expert specializing in **Hash Collisions**.
Your target is to break a solution that uses **Polynomial Rolling Hash** by generating two different strings with the same hash value (Collision).

\vspace{0.5em}
\noindent\textbf{\# Workflow}
\begin{enumerate}
    \setlength\itemsep{0em}
    \item \textbf{Identify Hashing}: Locate the hash calculation logic (usually \texttt{h = (h * B + c) \% M}).
    \item \textbf{Extract Parameters}:
    \begin{itemize}
        \item \textbf{Base (B)}: The multiplier (e.g., 31, 131, 13331, or a random value).
        \item \textbf{Modulus (M)}: The modulo (e.g., $10^9+7$, $10^9+9$).
        \item \textbf{Character Set (C)}: The valid alphabet (e.g., 'a'-'z').
        \item \textbf{Mapping (C)}: The character-to-integer conversion logic (e.g., \texttt{s[i] - 'a' + 1} or raw ASCII values).
        \item \textit{Note}: If the code uses \texttt{unsigned long long} without explicit modulo, $M = 2^{64}$.
    \end{itemize}
\end{enumerate}

\noindent\textbf{\# Input} \\
\textbf{Problem:} \texttt{\{problem\_description\}} \\
\textbf{Code:} \texttt{\{target\_code\}}

\vspace{0.5em}
\noindent\textbf{\# Output Format (JSON)}
\begin{lstlisting}[style=jsonstyle]
{
  "vulnerability_type": "Hash Collision",
  "hash_parameters": {
      "base": "List of Integer (e.g., [131, 13331])",
      "modulus": "List of Integer or expression (corresponding one-to-one with base, e.g., [998244353, 2^64])",
      "character set": "a,b,c,...,x,y,z",
      "mapping": "{'a':1,'b':2,...}"
  }
}
\end{lstlisting}
\end{promptbox}

\end{document}